\documentclass[a4paper,11pt]{article}
\usepackage{jheppub} % for details on the use of the package, please see the JINST-author-manual
\usepackage{lineno}
\usepackage{multirow, graphicx,amssymb,url,mathrsfs,amsmath,mathtools}
\usepackage{wrapfig,boxedminipage,setspace,subfigure,epsfig}
\usepackage{amsxtra,amstext,latexsym,dsfont,amsfonts}
\usepackage{color,eucal}
\usepackage[dvipsnames]{xcolor}
\usepackage{float}
\usepackage{slashed,comment}
\usepackage{tensor}
\usepackage{indentfirst}
\usepackage{setspace}

\newcommand{\dd}{\mathrm{d}}

\title{Holographic \textit{n}-partite Information in Hyperscaling Violating Geometry}

\author[a]{Xin-Xiang Ju,}
\author[a]{Teng-Zhou Lai,}
\author[a,b]{Ya-Wen Sun,}
\author[a]{and Yuan-Tai Wang}

\emailAdd{juxinxiang21@mails.ucas.ac.cn}
\emailAdd{laitengzhou20@mails.ucas.ac.cn}
\emailAdd{yawen.sun@ucas.ac.cn}
\emailAdd{wangyuantai19@mails.ucas.ac.cn}

\affiliation[a]{School of Physical Sciences, University of Chinese Academy of Sciences, Zhongguancun east road 80, Beijing 100190, China}
\affiliation[b]{Kavli Institute for Theoretical Sciences, University of Chinese Academy of Sciences, \\ Zhongguancun east road 80, Beijing 100049, China}

\abstract{The $n$-partite information (nI) is formulated as a measure of multi-partite entanglement. 
Field theory computation revealed that the sign of nI is indefinite for $n\geq 3$, while holographic studies conjectured a sign property that holographic nI is non-negative/non-positive for even/odd $n$, with tripartite information (TI, $n=3$) proved.
We investigate the aspects of nI with holographic duality in hyperscaling violating geometry. We confirm the conjectured sign property for strips of equal length with equal separation distance, and disprove this conjecture for $n>3$ with general configurations. 
Therefore, nI in field theories and holography exhibits compatibility except for $n=3$.
We also discuss other properties of holographic nI with analytic computation: the monotonicity, linearity, relation to hyperscaling violating parameters, temperature and UV cutoff effects, and the physical implications. 
It is doubtful that nI is an effective measure of entanglement considering the indefinite sign, non-monotonicity, and quasi-linearity of its holographic dual. In this respect, we propose constraints on the multi-partite entanglement measures.}

\begin{document}
\maketitle
\flushbottom

\section{Introduction}\label{1}

Quantum entanglement is a phenomenon in which two or more particles interact in such a way that the quantum state of each particle cannot be described independently, even when the particles are spatially separated \cite{sep-qt-entangle,Zou_2021}. The entanglement has been the significant subject of theoretical and experimental research in recent years, in which the entanglement measurement is a long-standing key problem.

Of the various measures of entanglement, entanglement entropy (EE) is one of the most representative, which quantifies the amount of quantum correlation between the degrees of freedom in a quantum mechanical system \cite{Callan:1994py,Horodecki:2009zz,Casini:2009sr,Calabrese:2009qy}. Consider in a local quantum field theory (QFT) a spatial region A and its complement $\Bar{A}$ on a constant time slice. The whole Hilbert space is partitioned to be $\mathcal{H}=\mathcal{H}_A\otimes\mathcal{H}_{\Bar{A}}$. Assume the density matrix for the whole system to be $\rho$, and then one defines the (geometric) EE for $A$ as the von Neumann entropy \cite{Bombelli:1986rw,Srednicki:1993im}
\begin{equation}\label{EE}
\begin{split}
\begin{aligned}
    S(A)&= -\mathrm{Tr}_A\; \rho_A\log\rho_A,\\
    \rho_A&=\mathrm{Tr}_{\Bar{A}}\; \rho,
\end{aligned}
\end{split}
\end{equation}
where $\rho_A$ is the reduced density matrix for $A$ obtained from the partial trace $\mathrm{Tr}_{\Bar{A}}$ over $\Bar{A}$. Similarly, one can define $S(\Bar{A})$, i.e. EE for $\Bar{A}$.

A summary of the properties of EE in field theories is as follows \cite{Calabrese:2004eu,Calabrese:2009ez,Casini:2009sr,Rangamani:2016dms}. First, EE is proportional to the entangling surface area with the local degrees of freedom entangled near the boundary surface. EE is UV divergent in QFTs, e.g. $S(A)\propto \frac{\mathrm{Area}(A)}{\epsilon^{d-2}}$ at leading order in $d$ dimensions, where $\epsilon$ is the UV cutoff. Given the small scale of the entangling region $A$, EE is a positive, increasing and concave function of the scale. If the whole system is in a pure state, the EEs for $A$ and $\Bar{A}$ are equal, i.e. $S(A)=S(\Bar{A})$. Whereas if the whole system is in a mixed state, $S(A)\neq S(\Bar{A})$ and EE is no longer a good entanglement measure because it mixes quantum and classical correlations. Besides, EE cannot capture all physical information of the field theory, e.g. EE is completely dependent on the central charge $c$ in two-dimensional conformal field theory (CFT). EE also satisfies some celebrated inequalities. For instance, sub-additivity $S(A)+S(B)\geq S(A\cup B)$ for two subsystems $A$ and $B$, Araki-Lieb inequality $S(A\cup B)\geq |S(A)-S(B)|$, and strong sub-additivity $S(A\cup B)+S(B\cup C)\geq S(A\cup B\cup C)+S(B)$ and $S(A\cup B)+S(B\cup C)\geq S(A)+S(C)$.

To overcome the drawbacks of EE, various entanglement measures have been proposed. For example, the mutual information (MI) measures the quantum and classical entanglement between two disjoint subsystems, which is defined as \cite{Groisman_2005,Casini:2008wt,Calabrese:2009ez,Alishahiha:2014jxa,Rangamani:2016dms,Agon:2022efa}
\begin{equation}\label{MI}
    I^{[2]}(A,B)=S(A)+S(B) - S(A\cup B).
\end{equation}
MI is finite if the two subsystems are disjoint, and diverges if the subsystems are contiguous. Computation shows that MI depends on the full operator content of the field theory \cite{Casini:2008wt,Calabrese:2009ez}. We know that $I^{[2]}\geq 0$ by using the sub-additivity of EE.

A natural generalization of MI to multi-partite subsystems is the $n$-partite information (nI). For $n=3$, the tripartite information (TI) is defined for three disjoint subsystems as \cite{Alishahiha:2014jxa,Rangamani:2016dms,Mirabi:2016elb,Agon:2022efa}
\begin{equation}\label{TI}
    I^{[3]}(A,B,C)=S(A)+S(B)+S(C)- \big( S(A\cup B)+S(A\cup C)+S(B\cup C) \big ) + S(A\cup B\cup C).
\end{equation}
TI is finite even when the subsystems share boundaries. Exceptionally, it will diverge if the common boundary is singular, e.g. conical singlularity in \cite{MohammadiMozaffar:2015wnx}. TI can be expressed in terms of MI, as one can check by definition that $I^{[3]}(A,B,C)=I^{[2]}(A,B)+I^{[2]}(A,C)-I^{[2]}(A,B\cup C)$. Therefore, TI measures the extensivity of MI. If TI is non-positive, MI is called monogamous. Whereas TI has no definite sign in a generic field theory, in contrast with the holographic counterpart (HTI) which we introduce in the following paragraph. 

For generic $n$, nI is defined for $n$ disjoint subsystems as \cite{Alishahiha:2014jxa,Mirabi:2016elb,Agon:2022efa}
\begin{equation}\label{nI}
    I^{[n]}(A_1,\cdots,A_n)=\sum_i S(A_i) - \sum_{i<j} S(A_i \cup A_j) + \cdots -(-1)^n S(\cup_i A_i).
\end{equation}
nI can be expressed in terms of MI, i.e. 
$$I^{[n]}(A_1,\cdots,A_n)=\sum_{i=2}^n I^{[2]}(A_1, A_i) - \sum_{i<j} I^{[2]}(A_1, A_i \cup A_j) + \cdots +(-1)^n I^{[2]}(A_1, \cup_{i=2}^n A_i),$$ from which we see that nI is finite. There is also a relation between $I^{[n]}$ and $I^{[n-1]}$, i.e. 
\begin{equation*}
\begin{split}
\begin{aligned}
    I^{[n]}(A_1,\cdots,A_n)&=I^{[n-1]}(A_1,\cdots,A_{n-2},A_{n-1})+I^{[n-1]}(A_1,\cdots,A_{n-2},A_{n}) \\
    &-I^{[n-1]}(A_1,\cdots,A_{n-2},A_{n-1}\cup A_n),
\end{aligned}
\end{split}
\end{equation*}
from which we see that the sign of $I^{[n]}$ determines the monogamy of $I^{[n-1]}$. However, $I^{[n]}$ has no definite sign for $n>2$ in field theories \cite{Alishahiha:2014jxa,Agon:2022efa}.

Computation of the entanglement measures is difficult in field theories where only limited analytic results were found. The holographic duality, exemplified by anti-de Sitter space/conformal field theory (AdS/CFT) correspondence, serves as a useful tool of the quantum entanglement research. Originally, AdS/CFT correspondence is a conjectured relationship between two theories: CFT in one lower-dimensional spacetime (the boundary) and a gravitational theory in a higher-dimensional spacetime (the bulk) \cite{Maldacena:1997re}. It states that the physics of the boundary theory can be described in terms of the bulk theory, and vice versa. It is a well-tested example of the more general holographic duality, i.e. gauge/gravity duality \cite{Maldacena:1997re,Gubser:1998bc,Witten:1998qj,maldacena2011gauge,Ramallo:2013bua,Hubeny:2014bla,Zaanen:2015oix}. In this way, quantum information in non-gravitational field theories has been studied holographically with gravitational computation.

In holography, the EE for the spacelike subregion $A$ on the boundary is evaluated by the Ryu-Takayanagi (RT) prescription \cite{Ryu:2006bv,Ryu:2006ef}. The dual quantity, that is, the holographic entanglement entropy (HEE) is at leading order proportional to the area of the extremal bulk surface (RT surface) $\Gamma_A$ homologous to $A$, i.e. $\partial\Gamma_A=\partial A$:
\begin{equation}\label{HEE}
    S(A)=\frac{\mathrm{Area}(\Gamma_A)}{4G_N} + \cdots,
\end{equation}
where $G_N$ is the bulk Newton constant and the ellipsis denotes quantum corrections. The Hubeny-Rangamani-Takayanagi (HRT) formula generalizes the RT prescription to time-dependent spacetimes, which was proposed in \cite{Hubeny:2007xt}.

\begin{figure}[htbp]
\centering
     \includegraphics[width=8.5cm]{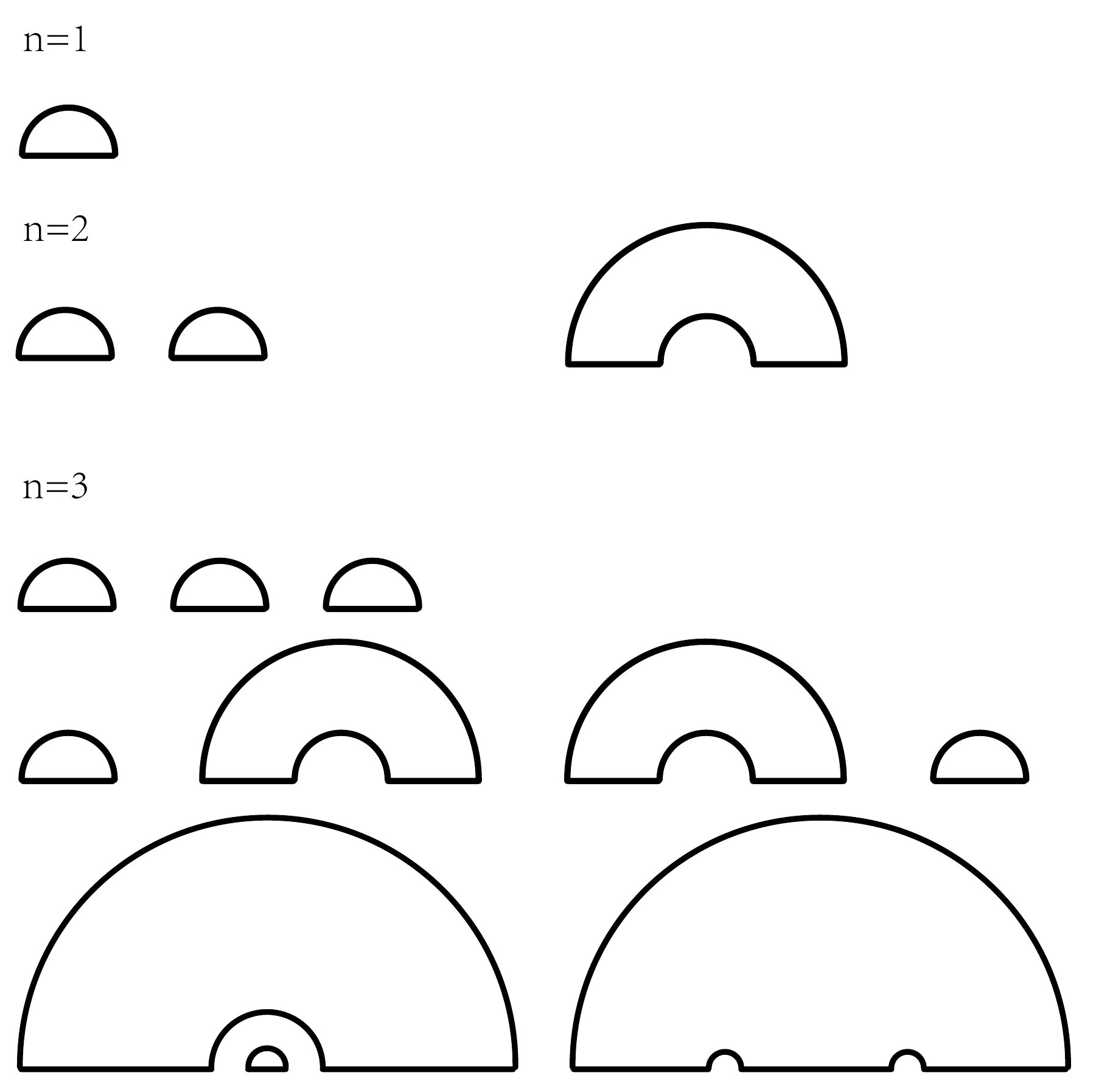}
 \caption{The configurations for HEE of one, two, and three parallel disjoint strips. The HEE of one, two, and three strips has one, two, and five possible configurations, respectively.}\label{fig_config}
\end{figure}

For disjoint regions $\cup_i A_i$, in principle one can perform a similar operation to find the minimum of all extremal surfaces homologous to $\cup_i A_i$ and apply the formula in eq. \ref{HEE} to compute $S(\cup_i A_i)$ \cite{Allais:2011ys}. Then the holographic $n$-partite information can be evaluated by definition, cf. eq. \ref{nI}. In one dimension, the configurations for the extremal surface of parallel strip-like regions $A_i$ were extensively studied \cite{Allais:2011ys,Hayden:2011ag,Alishahiha:2014jxa,Rangamani:2016dms,Mirabi:2016elb,Asadi:2018lzr,Mahapatra:2019uql}. For instance, $n=2$ yields two configurations: the connected configuration and disconnected configuration. Let the strip lengths be $l_i$ and the separation distance be $h$ ($h_i$ for more strips), and one would find that the connected configuration predominates at small $h/l$ and the disconnected predominates at large $h/l$, with a first-order phase transition at the critical distance. The two configurations indicate finite and zero entanglement respectively, which we will review in more detail in section \ref{3.2}. Similarly, $n=3$ yields five possible configurations \cite{Allais:2011ys}. The configurations for HEE of one, two, and three strips are displayed in figure \ref{fig_config}. Whereas the analytic result is difficult to find in two or higher dimensions. 

The relations and properties of nI were verified for its holographic dual, except that the holographic tripartite information (HTI) was found to be non-positive, or equivalently the holographic mutual information (HMI) is monogamous \cite{Hayden:2011ag}. Holographic nI has no definite sign for $n>3$ in general, with examples found in \cite{Hayden:2011ag,Mirabi:2016elb,Erdmenger:2017gdk}. However, most recent studies have found with analytic/numeric computation that holographic nI is non-negative/non-positive for even/odd $n$ in AdS/CFT, EMd/QCD, asymptotic flat spacetime/BRSFT, and Vaidya holography, etc. \cite{Alishahiha:2014jxa,MohammadiMozaffar:2015wnx,Mirabi:2016elb,Mahapatra:2019uql}. It is anticipated that this conjectured sign property helps examine the holographic duality, i.e. only the QFT with nI having a definite sign could have the dual gravitational description. The discrepancies between TI and HTI as well as holographic nI for different values of $n$ attract our attention. Besides, other interesting properties of holographic nI have not been formally discussed, e.g. its relation to the separation distance. Intuitively, we expect holographic nI to be a monotonic function of the distance, indicating an attenuating entanglement as the disjoint subregions separate. Remarkably, it can be graphically discovered in previous literature that the relation is non-monotonic for $n\geq 3$ even in AdS geometry \cite{Alishahiha:2014jxa,MohammadiMozaffar:2015wnx,Mirabi:2016elb}.

We also notice another multi-partite entanglement measure, i.e. the $n-$partite entanglement (nE), defined as \cite{Horodecki:2009zz,Alishahiha:2014jxa}
\begin{equation}\label{nE}
    J^{[n]}(A_1,\cdots,A_n)=\sum_i S(A_i) - S(\cup_i A_i).
\end{equation}
$J^{[n]}$ reduces to MI for $n=2$, is obviously non-negative, and attenuates to zero as the disjoint subregions separate.

We aim to examine the sign of holographic nI as well as the relations of holographic nI to the separation distance and other geometric parameters. In particular, we will study the relations between holographic nI and other geometric parameters in hyperscaling violating (HV) geometry since these calculations could be difficult without specifying the background geometry. This bulk geometry is dual to HV QFT on the boundary, which we will introduce in more detail in section \ref{3.1}. We will demonstrate that the holographic nI behaviors are distinct for $n=2$, $n=3$, and $n>3$, and attempt to provide physical interpretations. By using HV geometry, this work is thus motivated: (1) to study the conjectured sign property of holographic nI and the dual relation of nI and holographic nI, (2) to study other geometric properties of holographic nI previously neglected, (3) to detect the physical implications from these behaviors and provide constraints on the candidates of multi-partite entanglement measures, (4) to lay foundation for future studies, e.g. in higher dimensions and dynamical geometries.

This paper is organized as follows. Section \ref{2} introduces some previous results and conventions used for this paper; section \ref{3} is devoted to the computation in HV geometry with an emphasis on the effects of $n$, HV exponents, temperature, and UV cutoff; section \ref{4} is the discussion section.

% RT formula for HEE
% properties of HnI, and it is unclear that: the sign. Other intersting behaviors...
% motivation: the sign and the physics behind in holography; more properties relevant to the validity of HnI as a good entanglement measure; aspects of HV geometry; lay foundation for future works.
% This paper is organized as follows.
\section{Conventions and basics}\label{2}

In this section, we review some basic facts of holographic nI in general holography and establish the conventions that will be used later in this work. We consider the $n$-partite information of $n\; (1<n<\infty)$ parallel disjoint strips $A_i,\; i=1,\cdots,n$ lying on a fixed time-slice of the boundary so that the RT prescription can be applied on the corresponding bulk time-slice. We assume that the lengths of the strips are $l_i,\; i=1,\cdots,n$ \footnote{$l_i$ measures the length of the $i$-th strip in the $x_1$ direction. The infinite lengths in other spatial dimensions, all denoted as $L$, would only contribute a constant factor to the entanglement measures, which are therefore not of our interest.}, and that the distances in between are $h_i,\; i=1,\cdots,n-1$. 

%We make a few definitions which will be used in the following sections.

The number of generic configurations for HEE of $n$ parallel disjoint strips $S(\cup_i A_i)$, according to the RT proposal, is $(2n-1)!!$. Among these configurations some have intersecting point(s) on the bulk extremal curves, which we call \textit{intersecting configurations}; others do not, which we call \textit{non-intersecting configurations}. As proved in \cite{Allais:2011ys}, given $l_i$ and $h_i$, only one of the non-intersecting configurations can be the optimal. 

Mathematically, a union of some homologous bulk curves and the boundary strips constitute an \textit{undirected graph} \footnote{Graph theory was also applied to search for other inequalities HEE satisfies, cf. \cite{Bao:2015bfa}.}, \textit{connected} (with it being one connected component) or \textit{disconnected} (with more than one component). Thus only one non-intersecting configuration for $S(\cup_i A_i)$ amounts to a connected graph, which we call the \textit{connected configuration}, while all the others are disconnected graphs with maximally $n$ separately connected components, which we call the \textit{maximally disconnected configuration}. 

Next, we review three facts regarding holographi nI, which will be used later for the discussion and computation in the following sections.

The number $C(n)$ of non-intersecting configurations for HEE of $n$ parallel strips is the Catalan number $C_{n}=\frac{1}{n+1}\,\tbinom{2n}{n}=\frac{(2n)!}{(n+1)!n!}$ in combinatorics \cite{stanley_2015,Erdmenger:2017gdk}. For instance, $C(1)=1,\;C(2)=2,\;C(3)=5,\;C(4)=14$, etc.. One can check these results by enumeration with no difficulty.  

At large distances, i.e. $h_i/l_j\rightarrow\infty$, the configuration for HEE of $n$ parallel strips is the maximally disconnected configuration. One can further check with combinatorics that holographic nI vanishes in this case
\begin{equation}\label{In_large_h}
\begin{split}
\begin{aligned}
    I^{[n]}(A_1,\cdots,A_n)&=\sum_i S(A_i) - \sum_{i<j} S(A_i \cup A_j) + \cdots -(-1)^n S(\cup_i A_i) \\
    &=(C_{n-1}^0 - C_{n-1}^1 + \cdots - (-1)^n C_{n-1}^{n-1}) \sum_i S(A_i) \\
    &=0.
\end{aligned}
\end{split}
\end{equation}

At small distance, i.e. $h_i/l_j\rightarrow 0$, the configuration for HEE of $n$ parallel strips is the 
connected configuration. Moreover, one can check that by identifying $l_i\equiv l,\; h_i\equiv h$, holographic nI in this small distance limit reduces to \cite{Alishahiha:2014jxa}
\begin{equation}\label{In_small_h}
    I^{[n]}(A_1,\cdots,A_n) = (-1)^n \Big (2S((n-1)l+(n-2)h) - S(nl+(n-1)h) - S((n-2)l+(n-3)h) \Big) , 
\end{equation}
which implies that holographic nI is non-negative/non-positive for even/odd $n$ in this equally small distance limit since $S(l)$ is a concave function of $l$.

The sign of $I^{[3]}$ in holography, or equivalently the monogamy of $I^{[2]}$ was strictly checked in \cite{Hayden:2011ag} \footnote{The other inequalities that HEE satisfies have also been studied, cf. \cite{Headrick:2007km,Headrick:2013zda}.}. Nonetheless, \cite{Hayden:2011ag} stated in holography that $I^{[n]}$ does not seem to have a definite sign for $n>3$ in general. The counter-examples, e.g. in \cite{Hayden:2011ag,Mirabi:2016elb,Erdmenger:2017gdk}, are not negligible, which we will investigate in more detail in section \ref{4}. One will fail to verify the conjectured sign property by enumerating the configurations of HEE even with the simplification $l_i\equiv l,\; h_i\equiv h$, as we tested in the early stage of this work. Whereas in contrast, most recent studies have verified this sign property that holographic nI is non-negative/non-positive for even/odd $n$  \cite{Alishahiha:2014jxa,MohammadiMozaffar:2015wnx,Mirabi:2016elb,Mahapatra:2019uql}. 

%We will figure out with gravitational computation the reason why counter-examples of the sign property were scarcely detected.
We present our analytic computation of holographic nI in HV geometry in the next section.

\section{Computation in hyperscaling violating geometry}\label{3}

In this section, we check the sign of holographic nI and also study other properties, e.g. relations with other geometric parameters. It is not an easy task without specifying the background geometry, and therefore we study these properties in a large class of geometry, i.e. hyperscaling violating (HV) geometry. Our strategy is as follows. In HV geometry mainly characterized by the exponents $d_e$ and $z$, we examine the sign property of holographic nI stated above, and also discuss the monotonicity and linearity properties with $d_e, h, z$ and some other parameters by analytic computation. We focus on the role that $d_e$ and the strip number $n$ play in determining the holographic nI behaviors, and briefly analyze the temperature and cutoff effects characterized by $r_h$ and $r_c$ respectively. 

\subsection{Geometric setup}\label{3.1}

HV geometry is a large class of spacetimes including the AdS geometry, with the metric of the pure spacetime \cite{Charmousis:2010zz,Gouteraux:2011ce}
\begin{equation}\label{HVmetric}
\begin{split}
\begin{aligned}
\dd s^2 = r^{-\frac{2\,d_e}{d}} \left( -r^{-2(z-1)} \dd t^2 + \dd r^2 + \dd x_{i}^2 \right), \quad d_e=d-\theta ,
\end{aligned}
\end{split}
\end{equation}
where $i=1,2,\cdots,d$. The HV geometry, dual to the HV QFT living on the boundary, is characterized by two parameters, i.e. dynamical critical exponent $z$ and hyperscaling violation exponent $\theta$. Therefore, $d_e=d-\theta$ measures the effective spacetime dimension. This geometry reduces to AdS geometry if $(z,\theta)=(1,0)$, or equivalently $(z,d_e)=(1,d)$.

To extend to the finite temperature geometry, one can add an emblackening factor in the metric in eq. \ref{HVmetric} \cite{Dong:2012se,Jeong:2022jmp}:
\begin{equation}\label{HVmetric2}
\begin{split}
\begin{aligned}
\dd s^2 = r^{-\frac{2\,d_e}{d}} \left( -r^{-2(z-1)} \, f(r)\, \dd t^2 + \frac{\dd r^2}{f(r)} + \dd x_{i}^2 \right) , \quad f(r)=1-\left(\frac{r}{r_{h}}\right)^{d_e+z} ,
\end{aligned}
\end{split}
\end{equation}
where $r_h$ is the horizon radius. The Hawking temperature $T$ is given by 
\begin{equation}\label{HawkingT}
\begin{split}
\begin{aligned}
T=\frac{r_{h}^{1-z}\,|f'(r_h)|}{4\pi}=\frac{1}{4 \pi} \frac{|d_e+z|}{r_{h}^{z}}.
\end{aligned}
\end{split}
\end{equation}
One can find that the $z$ effect on holographic entanglement measurements appears only at finite temperature. With the null energy condition (NEC) and a positive specific heat condition, the physically relevant range for ($z, d_e$) is \cite{Dong:2012se,Jeong:2022jmp}
\begin{equation}\label{range}
\begin{split}
\begin{aligned}
d_e=0:& \quad z\leq0 \text{\quad or\quad} z\geq1 \,, \\ 
0<d_e<1:& \quad z \geq 2 - \frac{d_{e}}{d} \,, \\
d_e\geq1:& \quad 1 \leq d_e < d \text{\,\,\,and\,\,\,} z \geq 2-\frac{d_e}{d}   \text{\quad or \quad} d_e \geq d \text{\,\,\,and\,\,\,} z \geq 1 \,.
\end{aligned}
\end{split}
\end{equation}
Readers can refer to \cite{Dong:2012se,Jeong:2022jmp,Hoyos:2010at,Ogawa:2011bz,Huijse:2011ef} for a more complete analysis.

Another aspect we consider is the UV cutoff $r_c$ in the bulk spacetime, which has been widely studied in \cite{McGough:2016lol,Taylor:2018xcy,Hartman:2018tkw,Kraus:2018xrn,Cardy:2018sdv,Bonelli:2018kik,Gross:2019ach,Gross:2019uxi,Alishahiha:2019lng} and is conjectured to be dual to $T\Bar{T}$-like deformed QFT living on the boundary of the spacetime. The theory is well-known when it comes to AdS bulk spacetime ($(z,\theta)=(1,0)$), whose dual field theory is the $T\Bar{T}$ deformed CFT:
\begin{equation}
    \frac{\partial S(\lambda)}{\partial \lambda} = \int \dd^{d+1} x \sqrt{g} \, X(x) \,.
\end{equation}
$\lambda$ is the deformation parameter dual to $r_c$, and $X$ is the $T\bar{T}$ deformation operator, which are written as
\begin{equation}
\begin{split}
\begin{aligned}
    \lambda = \frac{4\pi G_{N}}{d+1}\,r_c^{d+1} \,,\qquad X = T^{\mu\nu}T_{\mu\nu} - \frac{1}{d}\left(T_\mu^\mu\right)^2 \,,
\end{aligned}
\end{split}
\end{equation}
where $T_{\mu\nu}$ is the stress tensor. The $T\bar{T}$ deformation has been extensively studied in terms of the correlation functions \cite{Kraus:2018xrn,Aharony:2018vux,He:2019ahx,Cardy:2019qao,He:2019vzf}, the entanglement entropy \cite{Dong:2012se,Donnelly:2018bef,Park:2018snf,Banerjee:2019ewu,Murdia:2019fax,Grieninger:2019zts,Jeong:2019ylz,Donnelly:2019pie,Khoeini-Moghaddam:2020ymm,Jeong:2022jmp}, and other entanglement measurements (MI, Renyi entropy, EWCS, etc.) \cite{Jeong:2019ylz,Khoeini-Moghaddam:2020ymm,BabaeiVelni:2023cge}. In our context of HV bulk geometry with the UV cutoff, the boundary field theory is conjectured as $T\Bar{T}$-like deformed HV QFT. Readers can refer to \cite{Alishahiha:2019lng} for more detail.

We aim to study the behavior of holographic nI in the HV geometry. To compute holographic nI, with the metric of pure HV spacetime in eq. \ref{HVmetric}, firstly, HEE in terms of the corresponding interval is evaluated by the RT prescription as \cite{Dong:2012se,Khoeini-Moghaddam:2020ymm,Jeong:2022jmp}
\begin{equation}\label{S0}
\begin{split}
\begin{aligned}
    S_0(l) &= l^{1-d_e}\big(\frac{1}{d_e-1}(\frac{l}{\epsilon})^{d_e-1} + \frac{2^{d_e-1}\Gamma_1^{d_e}}{1-d_e}\big),\; &\text{if}\; d_e\neq 1, \\
    \text{or}&= l^{1-d_e}\log(\frac{l}{\epsilon}),\; &\text{if}\; d_e = 1,
\end{aligned}
\end{split}
\end{equation}
where $\Gamma_1=\frac{\sqrt{\pi}\Gamma\big(\frac{d_e+1}{2d_e} \big)}{\Gamma\big(\frac{1}{2d_e} \big)}$ and we have dropped the prefactor $\frac{L^{d-1}}{2G_N}$ for simplicity. We emphasize that the spacetime dimension $d$ appears only in this prefactor, and that $d_e$ instead plays the role of the effective dimension in the HEE formula. One can take $d_e=d,\;z=1$ to recover the formula with $d$ for the AdS geometry. One can check that HEE in HV spacetime is an increasing function of $d_e$. \footnote{Here we use the dimensionful HEE, i.e. eq. \eqref{S0} without normalizing the length dimension, to study the $d_e$ effect. Remarkably, the dimensionless HEE, i.e. eq. \eqref{S0} normalized by the cutoff, is a decreasing function of $d_e$, which was graphically displayed in \cite{Khoeini-Moghaddam:2020ymm}.} As we expect, the entangling degrees of freedom in a pure state increase as the dimension becomes larger.

In the bulk of this section, we simplify the calculations by taking the strip lengths and the distances to be $l_i\equiv l,\; h_i\equiv h$. Then holographic nI in eq. \ref{nI} can be calculated with each term being the minimum of all configurations. We further set $l_i\equiv 1$ so that the quantities with a length dimension are implicitly normalized to be dimensionless. For example, the distance $h\equiv h/l$ denotes the ratio. We evaluate holographic nI and examine its relations to the geometric parameters for $n=2,\;3,\;4$ and generic $n$ separately in the following subsections.

\subsection{\textit{n}=2 and 3}\label{3.2}

\begin{figure}[htbp]
\centering
\includegraphics[width=10.5cm]{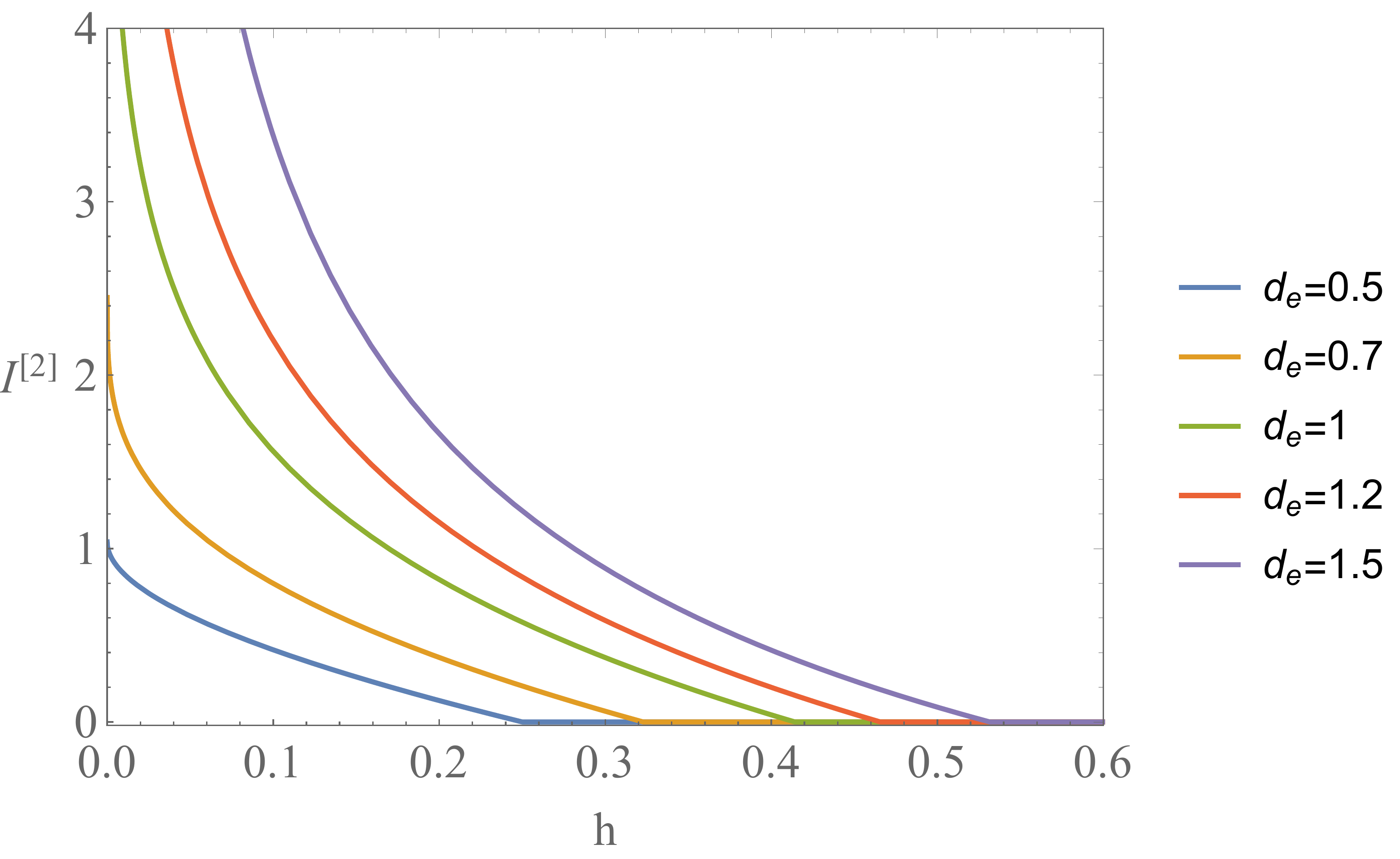}
 \caption{$I^{[2]}-h$ relation in HV spacetime for different discrete values of $d_e$. The strip lengths are set as $l_i\equiv 1$; all the separation distances are assumed to be $h_i\equiv h$.}\label{fig_I2}
\end{figure}

First, we briefly discuss the behavior of $I^{[2]}$. $S(A\cup B)$ has two kinds of minimal configurations, i.e. the disconnected and the connected. With the competition between the two configurations, it is easy to check that 
\begin{equation}\label{I2(A,B)}
\begin{split}
\begin{aligned}
    S(A\cup B) &= S(2l+h)+S(h) ,\; &\text{if}\; h<h_{c,2} \\
      \text{or}&= 2S(l),\; &\text{if}\; h>h_{c,2} \\
    \Rightarrow I^{[2]}(A,B) &= \max\{ 0,\; 2S(l)-S(h)-S(2l+h) \},
\end{aligned}
\end{split}
\end{equation}
where the critical distance is denoted as $h_{c,2}$. The $I^{[2]}-h$ relation is displayed in figure \ref{fig_I2}, in the range $d_e>0$. We see that $I^{[2]}(h)$ is a positive and monotonically decreasing function and has a first-order phase transition at $h_{c,2}$ due to the transition between the two configurations of $S(A\cup B)$. $I^{[2]}$ increases as $d_e$ increases, which indicates as we expected that the bipartite entanglement increases as the number of degrees of freedom increases. This behavior is the same as that found in \cite{Fischler:2012uv}. Note that $d_e$ plays the role of the effective dimension in HV geometry. The sign and other properties of $I^{[2]}$ were extensively discussed in \cite{Groisman_2005,Casini:2008wt,Calabrese:2009ez,Fischler:2012uv,Khoeini-Moghaddam:2020ymm,Alishahiha:2014jxa,Rangamani:2016dms,Agon:2022efa,Hayden:2011ag,Asadi:2018lzr,Mahapatra:2019uql,Mirabi:2016elb,Jain:2020rbb,Jain:2022hxl,Maulik:2022hty}. 

\begin{figure}[htbp]
\centering
     \includegraphics[width=10.5cm]{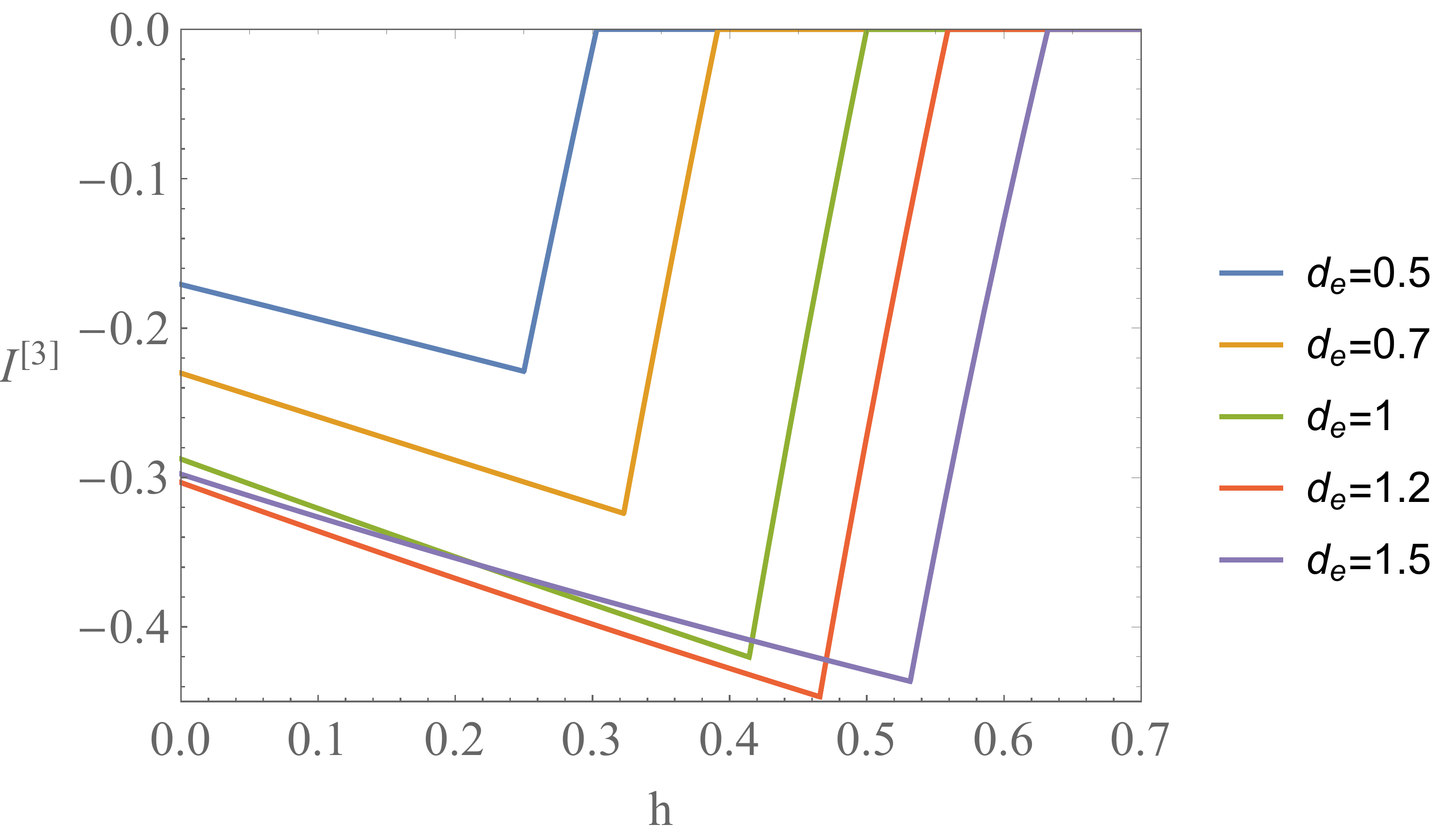}
 \caption{$I^{[3]}-h$ relation in HV spacetime for different discrete values of $d_e$. The strip lengths are set as $l_i\equiv 1$; all the separation distances are assumed to be $h_i\equiv h$. The first/second turning point in each curve corresponds to the value of $h_{c,2}$/$h_{c,3}$ in their respective case.}\label{fig_I3}
\end{figure}

Next, we turn to $I^{[3]}$. The last term, i.e. $S(A\cup B\cup C)$, only has two kinds of minimal configurations by identifying $l_i$ and $h_i$, i.e. the maximally disconnected and the connected, which occurs at the critical distance $h_{c,3}$. The configuration for $S(A\cup C)$ is always disconnected. As a consequence, one can check that \cite{Alishahiha:2014jxa,Mirabi:2016elb,Maulik:2022hty}
\begin{equation}\label{I3(A,B,C)}
\begin{split}
\begin{aligned}
    S(A\cup B\cup C) &= S(3l+2h)+2S(h) ,\; &\text{if}\; h<h_{c,3}, \\
            \text{or}&= 3S(l),\; &\text{if}\; h>h_{c,3}, \\
    S(A\cup C) &= 2S(l),\\
    \Rightarrow I^{[3]}(A,B,C) &= S(l)-2S(h+2l)+S(2h+3l),\; &\text{if}\; h<h_{c,2}, \\
                      \text{or}&= 2S(h)-3S(l)+S(2h+3l),\; &\text{if}\; h_{c,2}<h<h_{c,3}, \\
                      \text{or}&= 0,\; &\text{if}\; h>h_{c,3} .
\end{aligned}
\end{split}
\end{equation}
The $I^{[3]}-h$ relation is displayed in figure \ref{fig_I3}, in the range $d_e>0$. We see that $I^{[3]}(h)$ is non-positive and non-monotonic. We will check the sign and monotonicity in terms of $h$ for generic $n$ in section \ref{3.4}. In the full range of $d_e>0$, there are always two first-order phase transitions in each of the $I^{[3]}(h)$ curves, which correspond to the transition of the two $S(A\cup B)$ configurations and of the two $S(A\cup B\cup C)$ configurations, respectively. ${I^{[3]}}$ was computed in AdS geometry in \cite{Alishahiha:2014jxa,MohammadiMozaffar:2015wnx,Mirabi:2016elb}, and in asymptotically Lifshitz geometry in \cite{Maulik:2022hty}, where the non-monotonic ${I^{[3]}}-h$ relation and the two phase transitions can be graphically observed in \cite{MohammadiMozaffar:2015wnx,Mirabi:2016elb,Maulik:2022hty} and are consistent with our computation in the HV geometry.

When $h$ is small, we find that the dependence on $d_e$ is also non-monotonic. $|I^{[3]}|$ increases as $d_e$ increases for $d_e<1$ while the $d_e$ effect is weak for $d_e>1$. Generally speaking, ${I^{[3]}}(h)$ curves are stretched towards the positive $h$-axis by increasing $d_e$. Besides, at larger $h$, each ${I^{[3]}}(h)$ curve has a peak value. This peak value varies non-monotonically as $d_e$ increases in the range $d_e>1$. There are two competing underlying effects: similar to $I^{[2]}$, the tripartite entanglement increases as the degrees of freedom grow, while the alternating sum of the HEE terms in $I^{[3]}$ renders this effect non-monotonic.

\begin{figure}[htbp]
\centering
     \includegraphics[width=10.5cm]{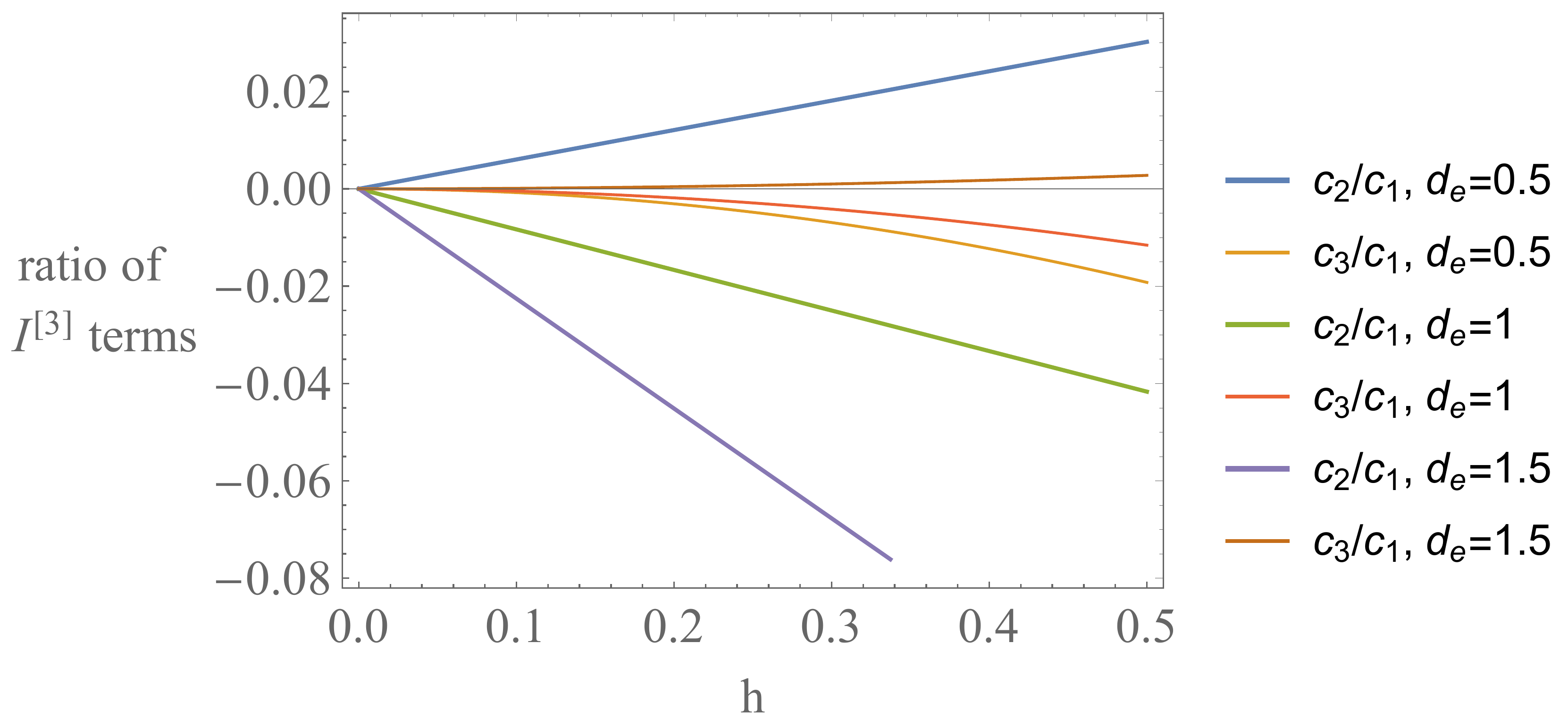}
 \caption{Ratios of $I^{[3]}(h)$ expansion terms at $h=0$ in HV spacetime for different discrete values of $d_e$. $c_i$ denotes the $i^{\text{th}}$ order term in the series expansion. The strip lengths are set as $l_i\equiv 1$; all the separation distances are assumed to be $h_i\equiv h$.}\label{fig_I3_coef}
\end{figure}

Moreover, counter-intuitively, the relation curves are quasi-linear in each segment. One can also observe the quasi-linearity graphically in the figures of \cite{MohammadiMozaffar:2015wnx,Mirabi:2016elb}. To study this phenomenon, we expand $I^{[3]}(h)$ at small $h$
\begin{equation}
\begin{split}
\begin{aligned}
    I^{[3]} &\propto \frac{6^{-d_e}(3*2^{d_e}-4*3^{d_e}+6^{d_e})}{1-d_e} + 2^{1-d_e}*3^{-d_e}(2^{d_e}-3^{d_e})h \\
    &-6^{-1-d_e}(2^{2+d_e}-3^{1+d_e})d_e h^2 + o(h^3),\; &\text{if}\; d_e\neq 1, \\
    \text{or}&\propto \log(\frac{3}{4}) - \frac{1}{3}h + \frac{1}{36}h^2 + o(h^3),\; &\text{if}\; d_e=1,
\end{aligned}
\end{split}
\end{equation}
and discover that even at larger $h$ exceeding the range $h<h_{c,2}$, the second and third-order terms are much smaller than the linear term, displayed in figure \ref{fig_I3_coef}. That is to say, the non-linearity is greatly weakened by the alternating sum of the HEE terms in $I^{[3]}$. In contrast, $I^{[2]}(h)$ is evidently non-linear, in that of the three terms, by fixing the strip length, $S(A)$ and $S(B)$ are invariants, and only $S(A \cup B)$ is a non-linear function of $h$, cf. eq. \eqref{I2(A,B)}. In analogy, one can study the $d_e$ effect and the quasi-linearity of $I^{[3]}(h)$ for $h>h_{c,2}$ segments. \footnote{$h_{c,2}$ is the critical distance for the phase transition of two strips, defined below eq. \eqref{I2(A,B)}. Therefore $h_{c,2}$ also appears as a phase transitioning point for more strips. In general, it is the first transitioning point, e.g. in eq. \eqref{I3(A,B,C)}.}

\subsection{\textit{n}=4}\label{3.3}

\begin{figure}[htbp]
\centering
     \includegraphics[width=10.5cm]{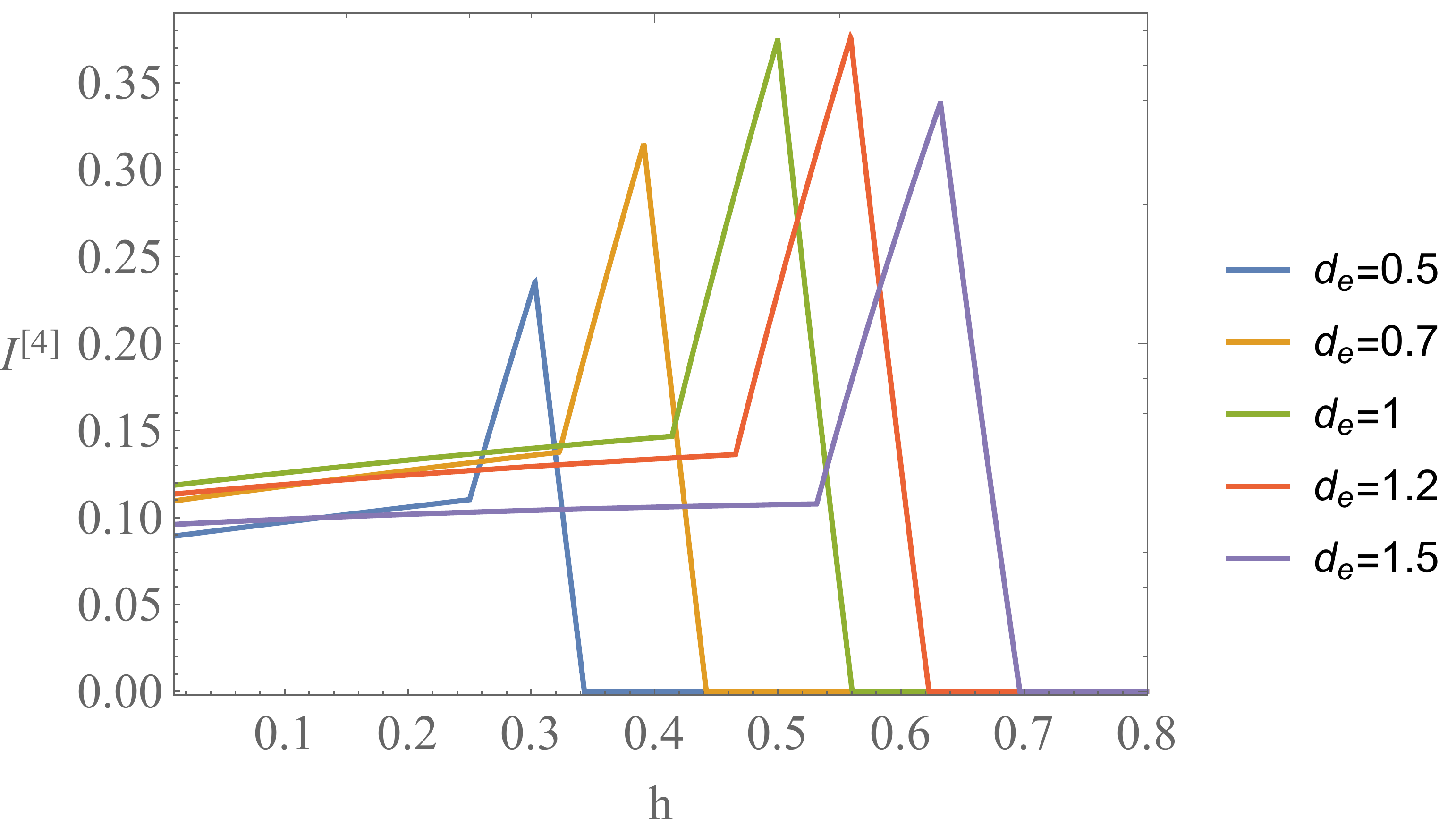}
 \caption{$I^{[4]}-h$ relation in HV spacetime for different discrete values of $d_e$. The strip lengths are set as $l_i\equiv 1$; all the separation distances are assumed as $h_i\equiv h$.}\label{fig_I4}
\end{figure}

Following the previous subsection, we study the behavior of $I^{[4]}$ in the HV geometry. ${I^{[4]}}$ was also computed in the AdS geometry in \cite{Alishahiha:2014jxa,MohammadiMozaffar:2015wnx,Mirabi:2016elb}. A complete analysis of the configurations of $S(\cup A_i)$ was made in \cite{Alishahiha:2014jxa,Mirabi:2016elb}, by assuming $l_i\equiv l,\; h_i\equiv h$. We compute $I^{[4]}$ by definition and display the graphical results in figure \ref{fig_I4}, in the range $d_e>0$. We see that $I^{[4]}(h)$ is a non-negative and non-monotonic function with three first-order phase transitions, which can also be detected in the figures of \cite{MohammadiMozaffar:2015wnx,Mirabi:2016elb}. \footnote{$I^{[n]}(h)$ has three first-order phase transitions for $n\geq 4$ by identifying the strip lengths and separation distances, cf. \cite{Alishahiha:2014jxa}.} 

Similar to the $n=3$ case, the $d_e$ effect is non-monotonic, and more precisely, ${I^{[4]}}(h)$ curves are likewise stretched towards the positive $h$-axis by increasing $d_e$. The peak value of $I^{[4]}$ varies non-monotonically as $d_e$ increases. Similarly, we consider with such kind of non-monotonicity that $I^{[4]}$ (as well as $I^{[n]}$ for larger $n$) is an ill-conditioned entanglement measure.

\begin{figure}[htbp]
\centering
     \includegraphics[width=10.5cm]{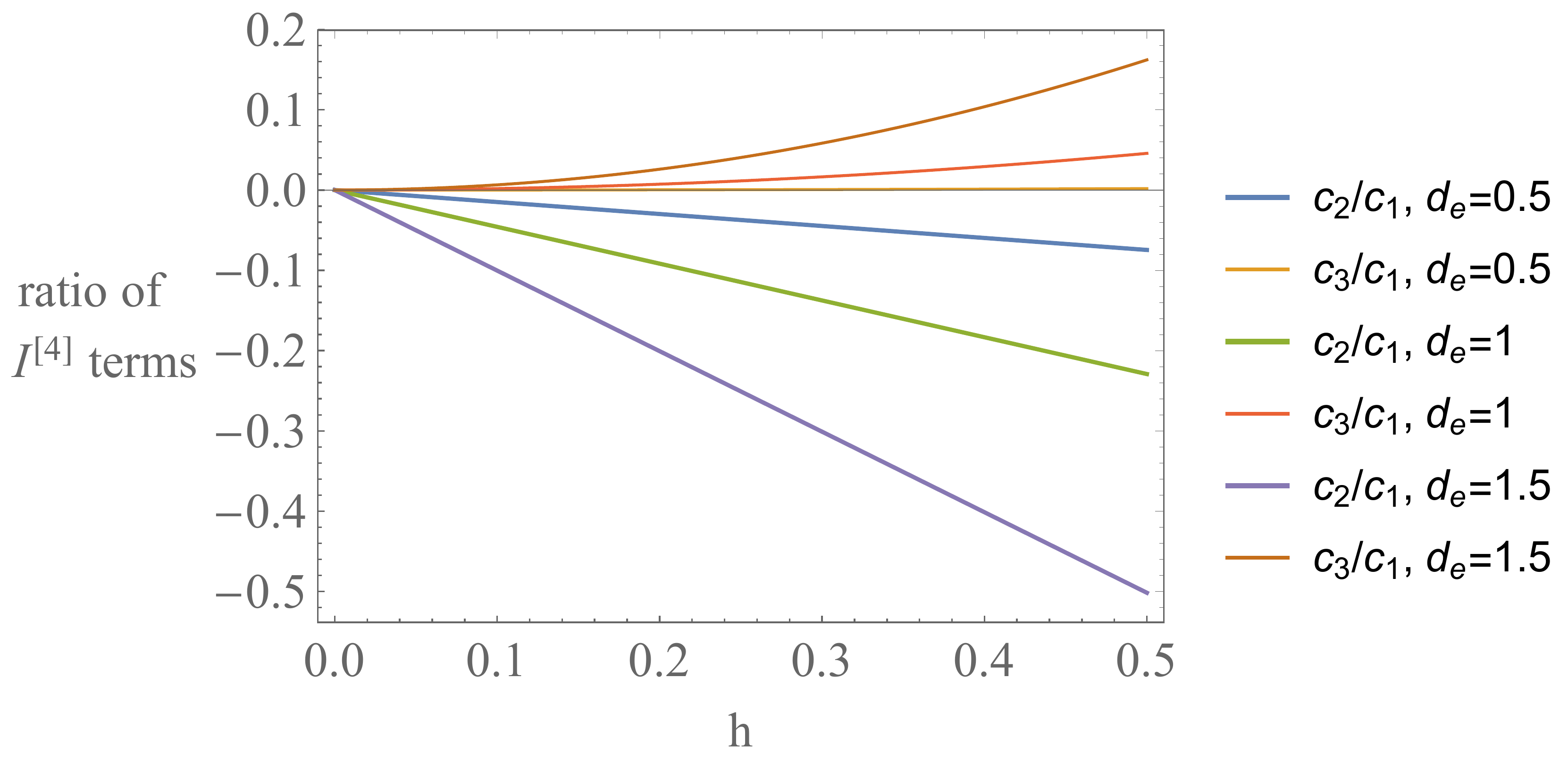}
 \caption{Ratios of $I^{[4]}(h)$ expansion terms at $h=0$ in HV spacetime for different discrete values of $d_e$. $c_i$ denotes $i^{\text{th}}$ order term. The strip lengths are set as $l_i\equiv 1$; all the separation distances are assumed as $h_i\equiv h$.}\label{fig_I4_coef}
\end{figure}

The $I^{[4]}(h)$ curves are also quasi-linear in the three segments (corresponding to non-vanishing $I^{[4]}$). This quasi-linearity can also be graphically verified in the figures of \cite{MohammadiMozaffar:2015wnx,Mirabi:2016elb}. In analogy, $I^{[4]}(h)$ at small $h$ has the series expansion
\begin{equation}
\begin{split}
\begin{aligned}
    I^{[4]} &\propto \frac{2^{1-2d_e}3^{-d_e}(3*2^{2d_e}-2*3^{d_e}-6^{d_e})}{1-d_e} + 2^{-2d_e}*3^{-d_e}(2^{2+2d_e}-3^{1+d_e}-6^{d_e})h \\
    &-2^{-3-2d_e}*3^{-1-d_e}(2^{5+2d_e}-3^{3+d_e}-6^{1+d_e})d_e h^2 + o(h^3),\; &\text{if}\; d_e\neq 1, \\
    \text{or}&\propto \log(\frac{9}{8}) + \frac{1}{12}h - \frac{11}{288}h^2 + o(h^3),\; &\text{if}\; d_e=1.
\end{aligned}
\end{split}
\end{equation}
We compare the ratios of the non-linear and linear terms in figure \ref{fig_I4_coef}. We find that at small $h$, $I^{[4]}(h)$ exhibits slightly stronger non-linearity compared to $I^{[3]}(h)$. In analogy, one can study the quasi-linearity of $I^{[4]}(h)$ at larger $h$.

\subsection{Generic \textit{n}}\label{3.4}

We have seen from figure \ref{fig_I3} and \ref{fig_I4} the behaviors of holographic nI for $n=3,4$, such as the sign, the phase transitions, non-monotonicity and quasi-linearity. In this subsection, we study two general properties of holographic nI ($n\geq 3$) in HV spacetime: the sign property and the monotonicity at small distance. The quasi-linear behavior can be probed for larger $n$ following the previous subsections. Most recent studies have found that holographic nI is non-negative/non-positive for even/odd $n$ in different holographic models \cite{Alishahiha:2014jxa,MohammadiMozaffar:2015wnx,Mirabi:2016elb,Mahapatra:2019uql}, and in particular, this sign property was analytically verified in the limit of small separation distance in the AdS geometry for equal strip lengths $l_i$ and distances $h_i$ \cite{Alishahiha:2014jxa}.

By identifying $l_i$ to $l$ and $h_i$ 
 to $h$ respectively, holographic nI at small distance $h$ can be described by eq. \ref{In_small_h}. Insert the HEE behavior in eq. \ref{S0} and we evaluate $I^{[n]}(h\rightarrow 0)$, i.e.
\begin{equation}
\begin{split}
\begin{aligned}
    I^{[n]} &= \frac{(-1)^{1+n}2^{-1+d_e}\big( n^{1-d_e}+(n-2)^{1-d_e}-2(n-1)^{1-d_e} \big)\Gamma_1^{d_e}}{1-d_e} l^{1-d_e} + \cdots,\; &\text{if}\; d_e\neq 1, \\
    \text{or}&= (-1)^n\log\big( \frac{(n-1)^2}{n(n-2)} \big)+ \cdots,\; &\text{if}\; d_e= 1, \\
\end{aligned}
\end{split}
\end{equation}
where the ellipsis denotes the higher order terms of $h$. Thus the sign is simply determined by $(-1)^n$ because the multiplication of $d_e$ terms always has a positive sign. Therefore, we have verified the sign property in HV spacetime that holographic nI for identical strips separated by equally small distances is positive/negative for even/odd $n$. In fact, the holographic nI formula in eq. \ref{In_small_h} implies the sign property at small distance as long as $S(l)$ is a concave function, regardless of the spacetime geometry. This is the reason why this sign property was verified broadly in previous literature \cite{Alishahiha:2014jxa,MohammadiMozaffar:2015wnx,Mirabi:2016elb,Mahapatra:2019uql}. 

One can further check the relationship between holographic nI and HEE behaviors, e.g. monotonicity of holographic nI, by taking the derivative of $h$ on both sides of eq. \ref{In_small_h}. We focus on HV background spacetime and calculate the first-order derivative of $I^{[n]}(h)$ at $h\rightarrow 0$, i.e.
\begin{equation}
\begin{split}
\begin{aligned}
    {I^{[n]}}'(0) &\propto (-1)^{n}\big( -(n-2)^{d_e}(n-1)^{1+de}+2(n-2)^{1+d_e}n^{d_e}-(n-3)((n-1)n)^{d_e} \big),\; &\text{if}\; d_e\neq 1, \\
    \text{or}&\propto \frac{2(-1)^n}{n(n-1)(n-2)},\; &\text{if}\; d_e= 1.
\end{aligned}
\end{split}
\end{equation}
Obviously, when $d_e=1$, the sign of ${I^{[n]}}'(0)$, or equivalently the monotonicity of $I^{[n]}(h)$ is determined by $(-1)^n$. When $d_e\neq 1$, we calculate ${I^{[n]}}'(0)$ by fixing integer $n$, e.g.
\begin{equation}
\begin{split}
\begin{aligned}
    {I^{[3]}}'(0) &\propto 2^{d_e}-3^{d_e},\\ 
    {I^{[4]}}'(0) &\propto -3^{1+d_e}+4^{1+d_e}-6^{d_e}.
\end{aligned}
\end{split}
\end{equation}
We find a salient feature at small $h$, summarized as follows. ${I^{[3]}}(h)$ is a decreasing function for arbitrary $d_e$. ${I^{[4]}}(h)$ is an in/decreasing function for $d_e<d_{e_c,4}/d_e>d_{e_c,4}$ with critical $d_{e_c,4}\sim 2.1$; ${I^{[5]}}(h)$ is a de/increasing function for $d_e<d_{e_c,5}/d_e>d_{e_c,5}$ with critical $d_{e_c,5}\sim 1.7$; this switching property of monotonicity universally holds according to the parity of $n$ for $n\geq 4$. Specifically, $d_{e_c}$ decreases as $n$ increases, with $d_{e_c}\rightarrow 1$ in the large $n$ limit. 

\begin{figure}[htbp]
\centering
     \includegraphics[width=8.5cm]{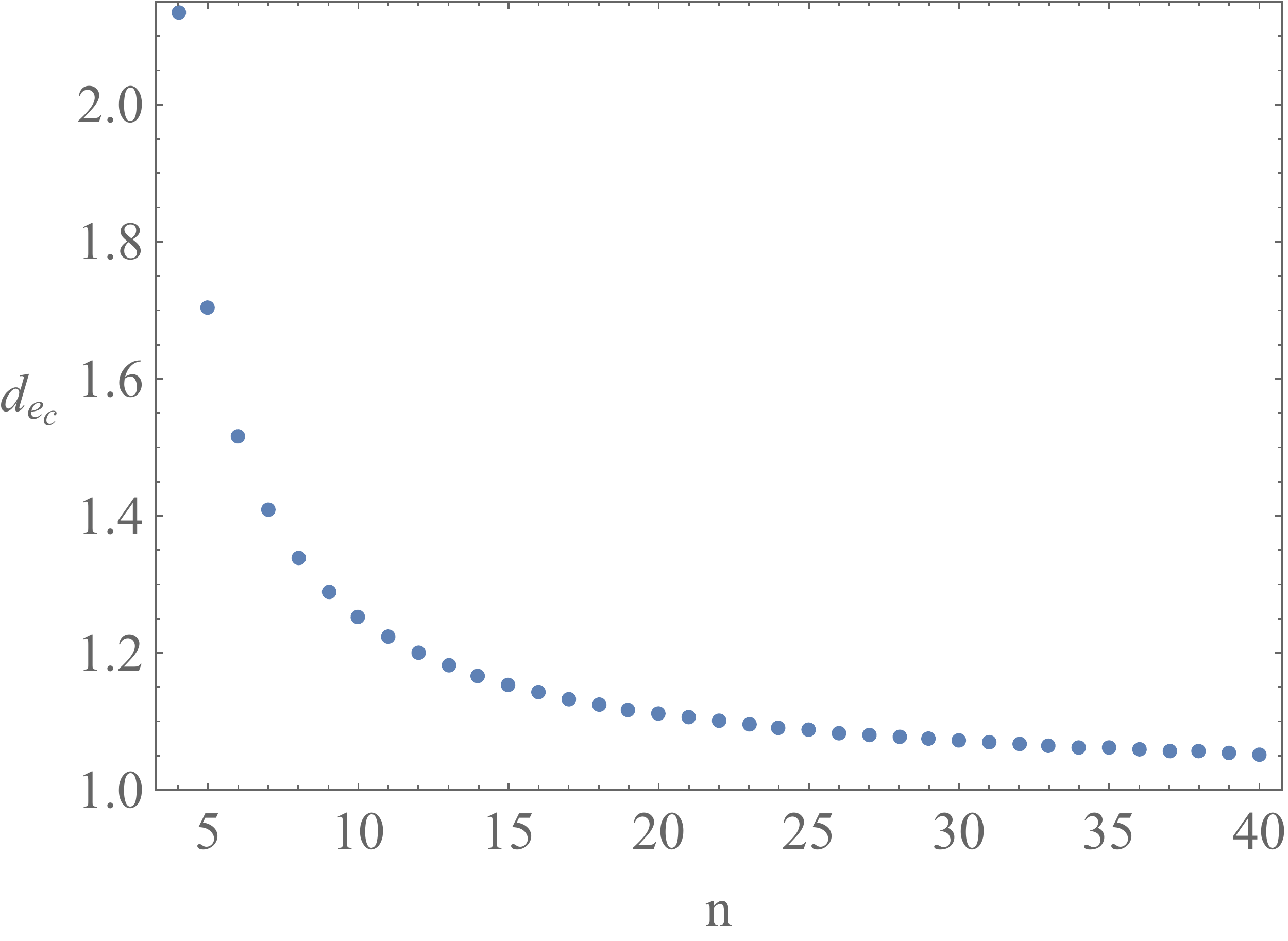}
 \caption{Relation of critical $d_e$ and $n$ in the HV spacetime. For $d_e$ smaller/larger than the critical $d_e$, the monotonicity of ${I^{[n]}}(h)$ at $h\rightarrow 0$ is the opposite. The strip lengths are set as $l_i\equiv 1$; all the separation distances are assumed as $h_i\equiv h$.}\label{fig_n_dec}
\end{figure}

\begin{figure}[htbp]
\centering
     \includegraphics[width=10.5cm]{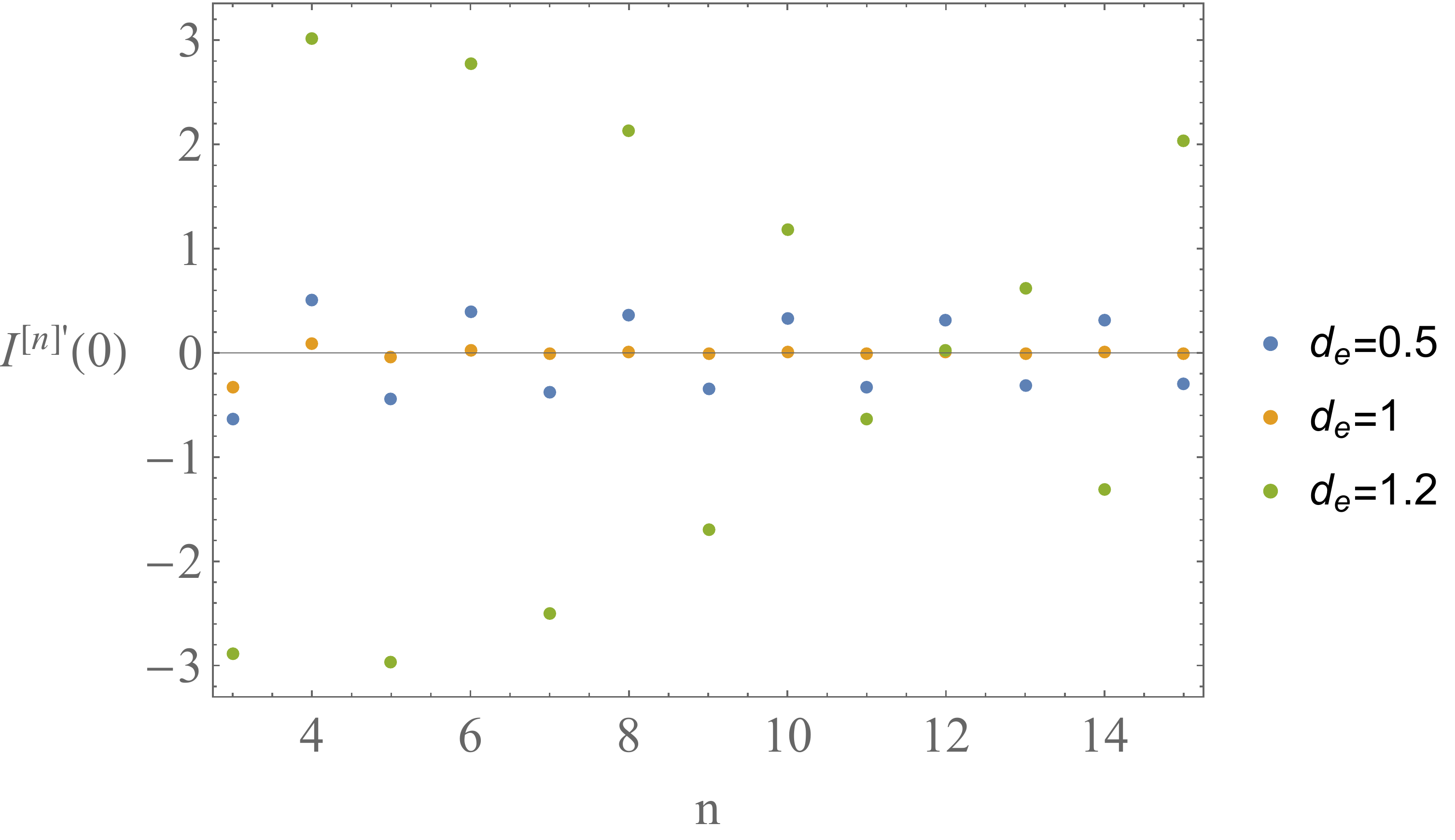}
 \caption{Values of ${I^{[n]}}'(0)$ in HV spacetime for different discrete values of $n$ and $d_e$. The strip lengths are set as $l_i\equiv 1$; all the separation distances are assumed as $h_i\equiv h$.}\label{fig_n_I'}
\end{figure}

Therefore, the $d_e=1$ case is not exceptional. Considering the full range of $d_e$, we put it another way. For $d_e\leq 1$, ${I^{[n]}}(h)$ is an in/decreasing function for even/odd $n$. For $d_e>1$, this feature holds for small $n$ but becomes the opposite for large $n$, where the critical $n_c$ decreases as $d_e$ increases, satisfying $3<n_c(d_e\rightarrow\infty)<4$. We display the critical $d_e$ by fixing $n$ in figure \ref{fig_n_dec}, and display the value of ${I^{[n]}}'(0)$ by fixing $n$ and $d_e$ in figure \ref{fig_n_I'}. 
Our discussion on the monotonicity depends on spacetime geometry, which in our case is the HV spacetime. Our analytics are highly consistent with the graphical results for pure AdS and small AdS black hole in \cite{Alishahiha:2014jxa,MohammadiMozaffar:2015wnx,Mirabi:2016elb}. We also notice that the temperature effect would reverse the monotonicity, which is observable in the numerical results for large AdS black hole \cite{Mirabi:2016elb}. 

In short, we have verified that holographic nI is non-negative/non-positive for even/odd $n$ in HV spacetime for strips of equal length and equally small separation distance, and that the relation of holographic nI and the distance is generally non-monotonic. Whereas we will quickly see that the sign property does not hold for $n\geq 4$ with generic configurations, which we discuss in section \ref{4}. Therefore, we do not consider that holographic nI is a physically well-defined entanglement measure. Moreover, we find in the HV geometry (including the thermal case in the next subsection) that $I^{[n\geq 3]}$ does not diverge in the small distance limit while $I^{[2]}$ could diverge, e.g. it is finite if $d_e<1$, and divergent if $d_e\geq 1$. In fact, this property can be examined by eq. \ref{In_small_h}, consistent with the general discussion in \cite{Hayden:2011ag} and the analytic computation in \cite{Khoeini-Moghaddam:2020ymm}.

\subsection{Temperature and cutoff effects}\label{3.5}

As an extension of the pure HV spacetime, finally we consider two deformations separately: small black hole (large inverse radius $r_h$) dual to the thermal HV QFT at low temperature $T$, and small UV cutoff $r_c$ dual to the conjectured $T\Bar{T}$-like deformed QFT. We expect that the temperature and cutoff have universal behaviors in different kinds of background spacetimes. For $n=2$, the temperature effect was studied in \cite{Fischler:2012uv}, and the cutoff effect was studied in \cite{Khoeini-Moghaddam:2020ymm}.

For the small black hole case, we assume that $l,h<<r_h$. HEE in terms of the corresponding interval with the leading order 1/$r_h$ correction is evaluated as \cite{Jeong:2022jmp}
\begin{equation}\label{S_T}
\begin{split}
\begin{aligned}
    S(l)&=S_0(l)+\Delta S_T(l),\\
    \Delta S_T(l)&=\frac{d_e \Gamma_2}{2^{2+z}(1+z)(1+d_e+z)\Gamma_1^{1+z}} (\frac{l}{r_h})^{d_e+z} l^{1-d_e},
\end{aligned}
\end{split}
\end{equation}
where $S_0(l)$ is the leading order vacuum contribution defined in eq. \ref{S0}, and $\Delta S_T(l)$ is the correction term where $\Gamma_2 = \frac{\sqrt{\pi}\Gamma(1+\frac{1+z}{2d_e})}{\Gamma(\frac{1}{2}+\frac{1+z}{2d_e})}$. Obviously, HEE increases as temperature increases in this zero cutoff limit.

\begin{figure}[htbp]
\centering
     \includegraphics[width=10.5cm]{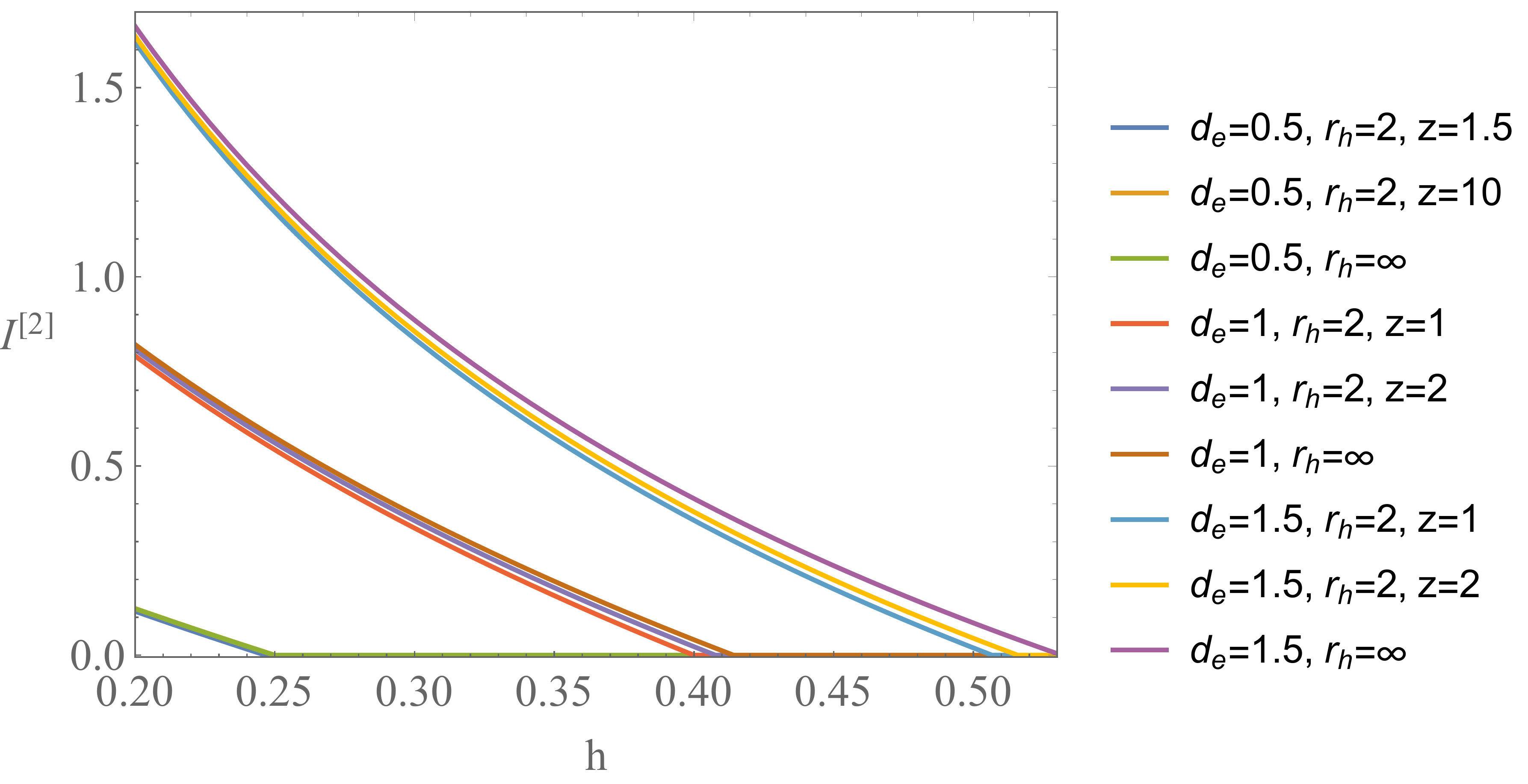}
 \caption{$I^{[2]}-h$ relation for different discrete values of $d_e$, $r_h$, and $z$. $r_h=\infty$ indicates zero temperature and finite $r_h$ indicates low temperature, corresponding to the HV spacetime and the small black hole respectively. The strip lengths are set as $l_i\equiv 1$; all the separation distances are assumed as $h_i\equiv h$.}\label{fig_I2_T}
\end{figure}

\begin{figure}[htbp]
\centering
     \includegraphics[width=10.5cm]{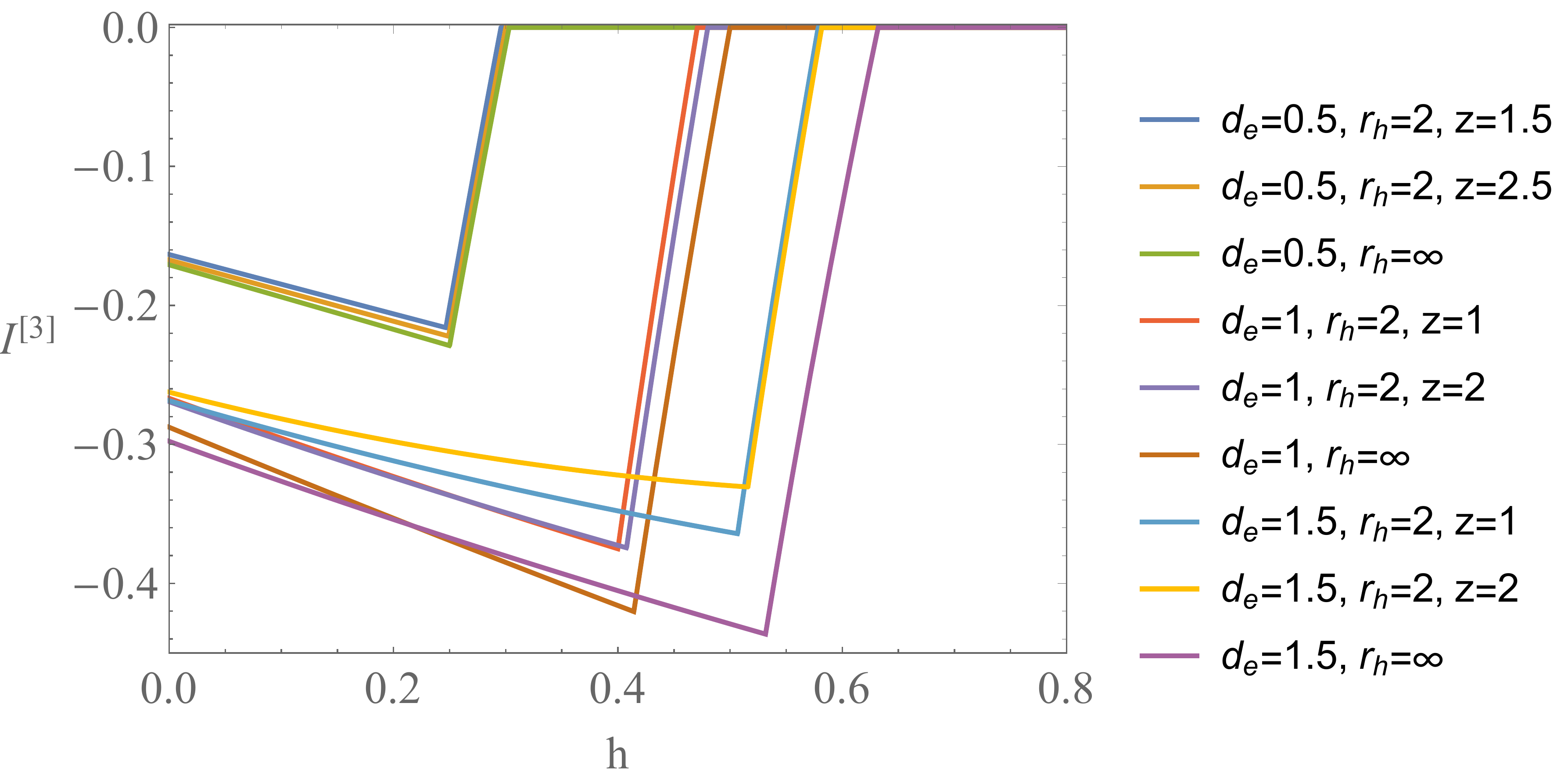}
 \caption{$I^{[3]}-h$ relation for different discrete values of $d_e$, $r_h$, and $z$. $r_h=\infty$ indicates zero temperature and finite $r_h$ indicates low temperature, corresponding to HV spacetime and small black hole respectively. The strip lengths are set as $l_i\equiv 1$; all the separation distances are assumed as $h_i\equiv h$.}\label{fig_I3_T}
\end{figure}

\begin{figure}[htbp]
\centering
     \includegraphics[width=10.5cm]{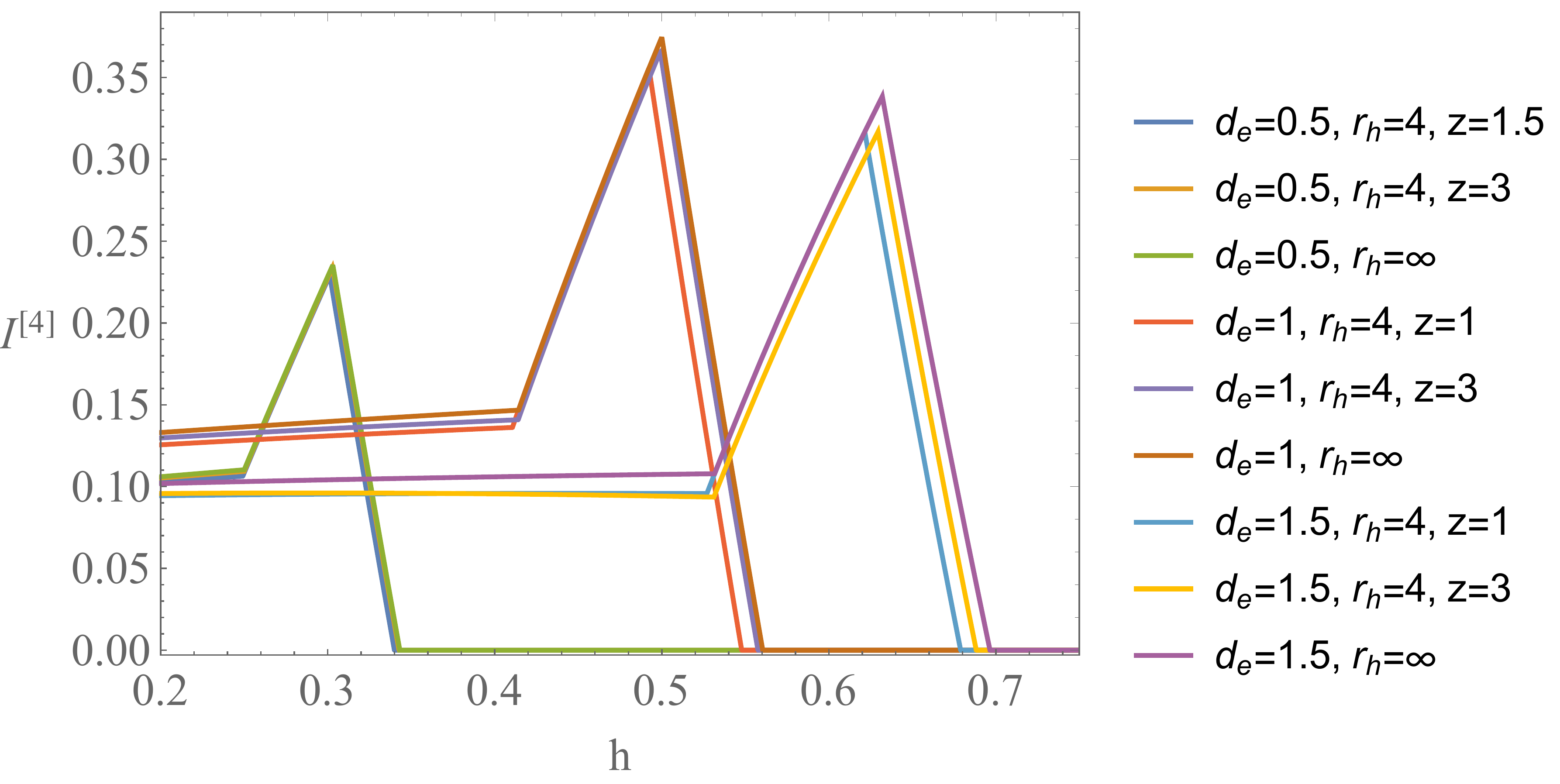}
 \caption{$I^{[4]}-h$ relation for different discrete values of $d_e$, $r_h$, and $z$. $r_h=\infty$ indicates zero temperature and finite $r_h$ indicates low temperature, corresponding to HV spacetime and small black hole respectively. The strip lengths are set as $l_i\equiv 1$; all the separation distances are assumed as $h_i\equiv h$.}\label{fig_I4_T}
\end{figure}

We evaluate holographic nI by definition, cf. eq. \ref{nI} and display the $I^{[n]}-h$ relation for $n=2,3,4$ in figure \ref{fig_I2_T},\ref{fig_I3_T},\ref{fig_I4_T}, in the range $d_e>0$. The $r_h$ and $z$ effects are manifest. In the full range $d_e>0$, $I^{[2]}$ decreases as the temperature increases, and increases as $z$ increases at finite temperature. $|I^{[3]}|$ and $I^{[4]}$ also decreases as the temperature increases, while the $z$ effect is non-monotonic, depending on $n$ and $d_e$. A possible explanation for the temperature effect is that the entangled degrees of freedom become free and disentangled. \cite{Fischler:2012uv} also suggested the possible relationship of temperature and disentanglement from the field theory side. The $z$ effect can be traced back to the scaling behavior of $z$ in HEE, cf. eq. \eqref{S_T}. The $I^{[2]}$-temperature relation we observe is consistent with \cite{Fischler:2012uv}.

For the small cutoff correction, we assume that $r_c<<l,h$. HEE with the leading order $r_c$ correction is evaluated as \cite{Khoeini-Moghaddam:2020ymm,Jeong:2022jmp}
\begin{equation}\label{S_rc}
\begin{split}
\begin{aligned}
    S(l)&=S_0(l)+\Delta S_c(l),\\
    \Delta S_c(l)&=\frac{(2\Gamma_1)^{2d_e}}{2(d_e+1)} (\frac{r_c}{l})^{d_e+1} l^{1-d_e},
\end{aligned}
\end{split}
\end{equation}
where we use $S_0(l)=\log\big(\frac{l}{r_c}\big)$ when $d_e=1$, referring to \cite{Khoeini-Moghaddam:2020ymm}, and $\Delta S_c(l)$ is the correction term. One can further consider higher order corrections with no difficulty. One can check that HEE is a decreasing function of $r_c$ \cite{Khoeini-Moghaddam:2020ymm,Jeong:2022jmp}.

\begin{figure}[htbp]
\centering
     \includegraphics[width=10.5cm]{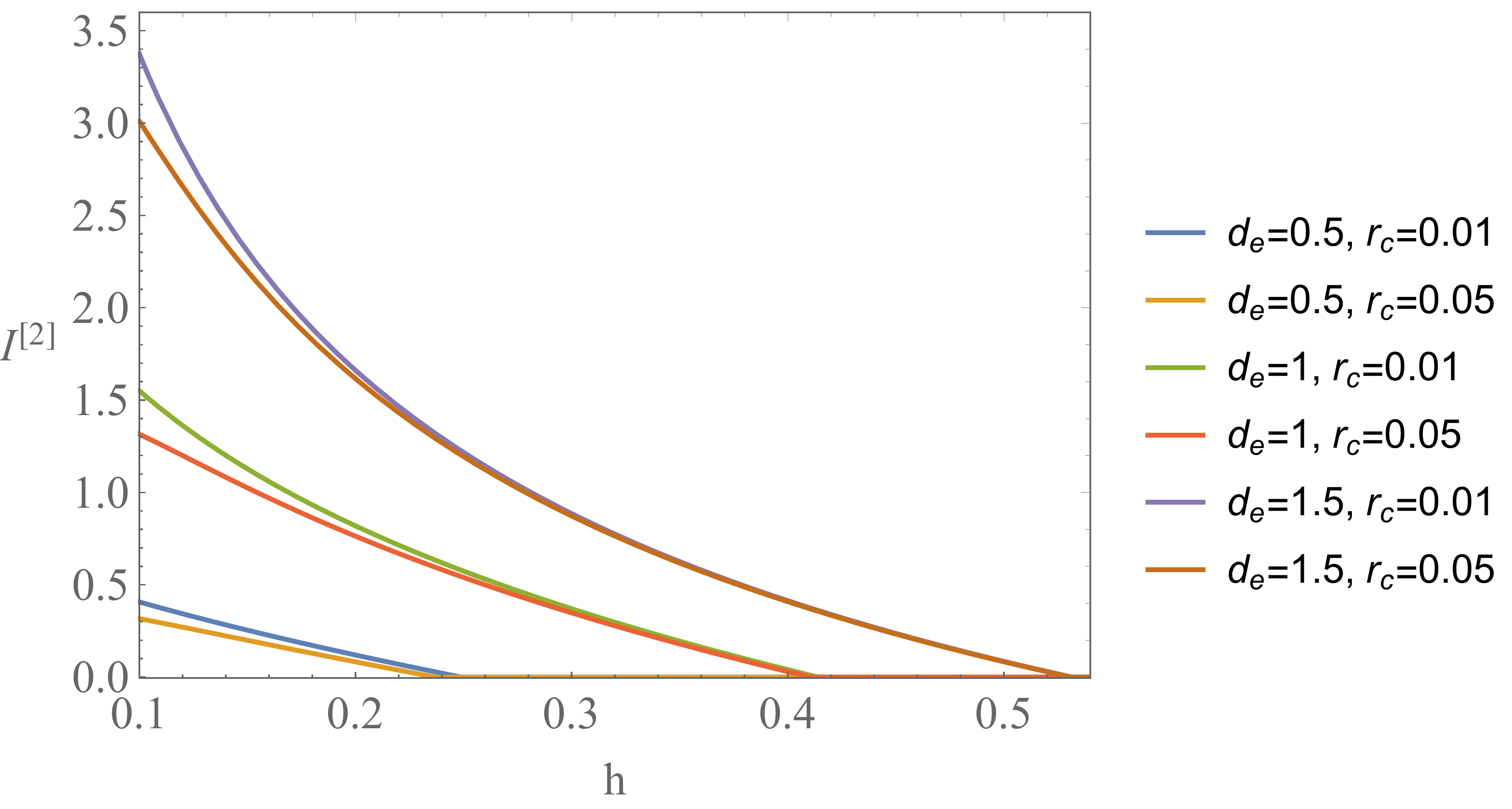}
 \caption{$I^{[2]}-h$ relation in HV geometries for different discrete values of $d_e$ and small cutoff $r_c$. $r_c\rightarrow 0$ indicates HV spacetime. The strip lengths are set as $l_i\equiv 1$; all the separation distances are assumed as $h_i\equiv h$.}\label{fig_I2_rc}
\end{figure}

\begin{figure}[htbp]
\centering
     \includegraphics[width=10.5cm]{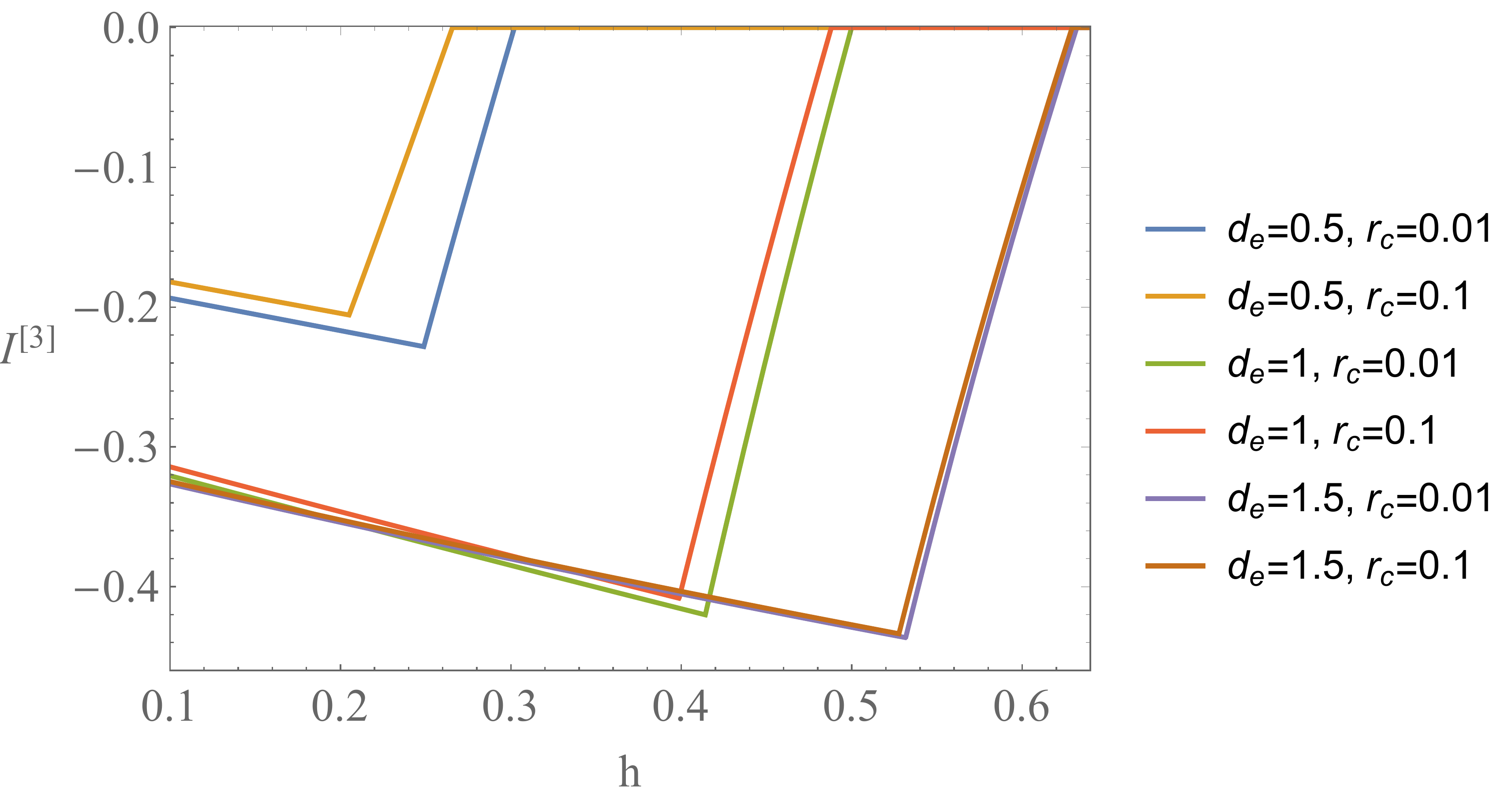}
 \caption{$I^{[3]}-h$ relation in HV geometries for different discrete values of $d_e$ and small cutoff $r_c$. $r_c\rightarrow 0$ indicates HV spacetime. The strip lengths are set as $l_i\equiv 1$; all the separation distances are assumed as $h_i\equiv h$.}\label{fig_I3_rc}
\end{figure}

\begin{figure}[htbp]
\centering
     \includegraphics[width=10.5cm]{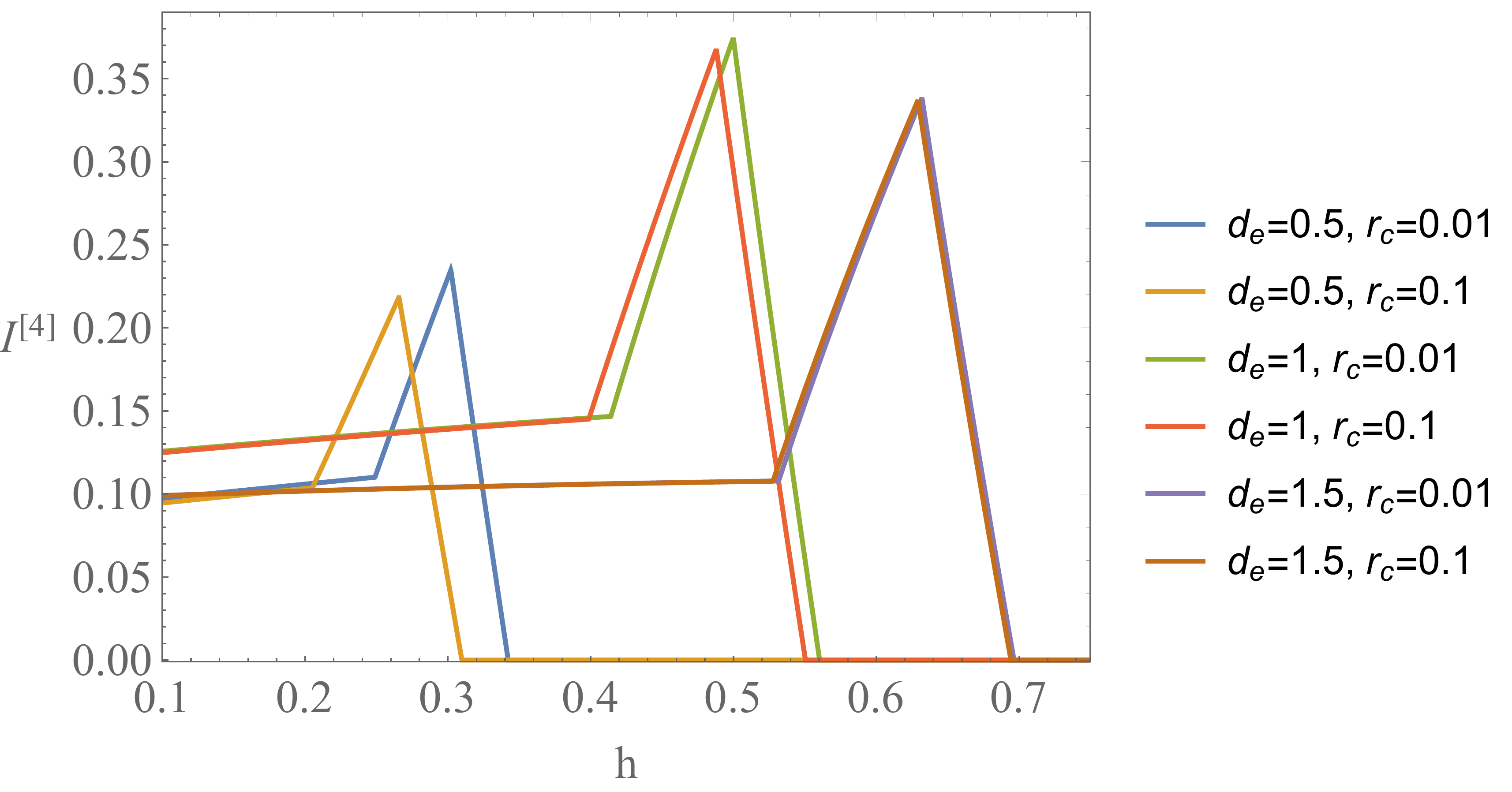}
 \caption{$I^{[4]}-h$ relation in HV geometries for different discrete values of $d_e$ and small cutoff $r_c$. $r_c\rightarrow 0$ indicates HV spacetime. The strip lengths are set as $l_i\equiv 1$; all the separation distances are assumed as $h_i\equiv h$.}\label{fig_I4_rc}
\end{figure}

We display the $I^{[n]}-h$ relation for $n=2,3,4$ in figure \ref{fig_I2_rc},\ref{fig_I3_rc},\ref{fig_I4_rc}, in the range $d_e>0$. The $r_c$ effect is clearly detected. In the full range $d_e>0$, $I^{[2]}$ decreases as $r_c$ increases. The cutoff could have an effect of diminishing the bipartite entanglement. For larger $n$, the $r_c$ effect is non-monotonic: roughly speaking, the $I^{[n]}(h)$ curves are stretched towards the negative $h$-axis by increasing $r_c$. We interpret this non-monotonous relation as another reflection that $I^{[n]}$ may be unphysical. Our result for $I^{[2]}$ is consistent with \cite{Khoeini-Moghaddam:2020ymm}.

\section{Discussion}\label{4}

\begin{figure}[htbp]
\centering
     \includegraphics[width=10.5cm]{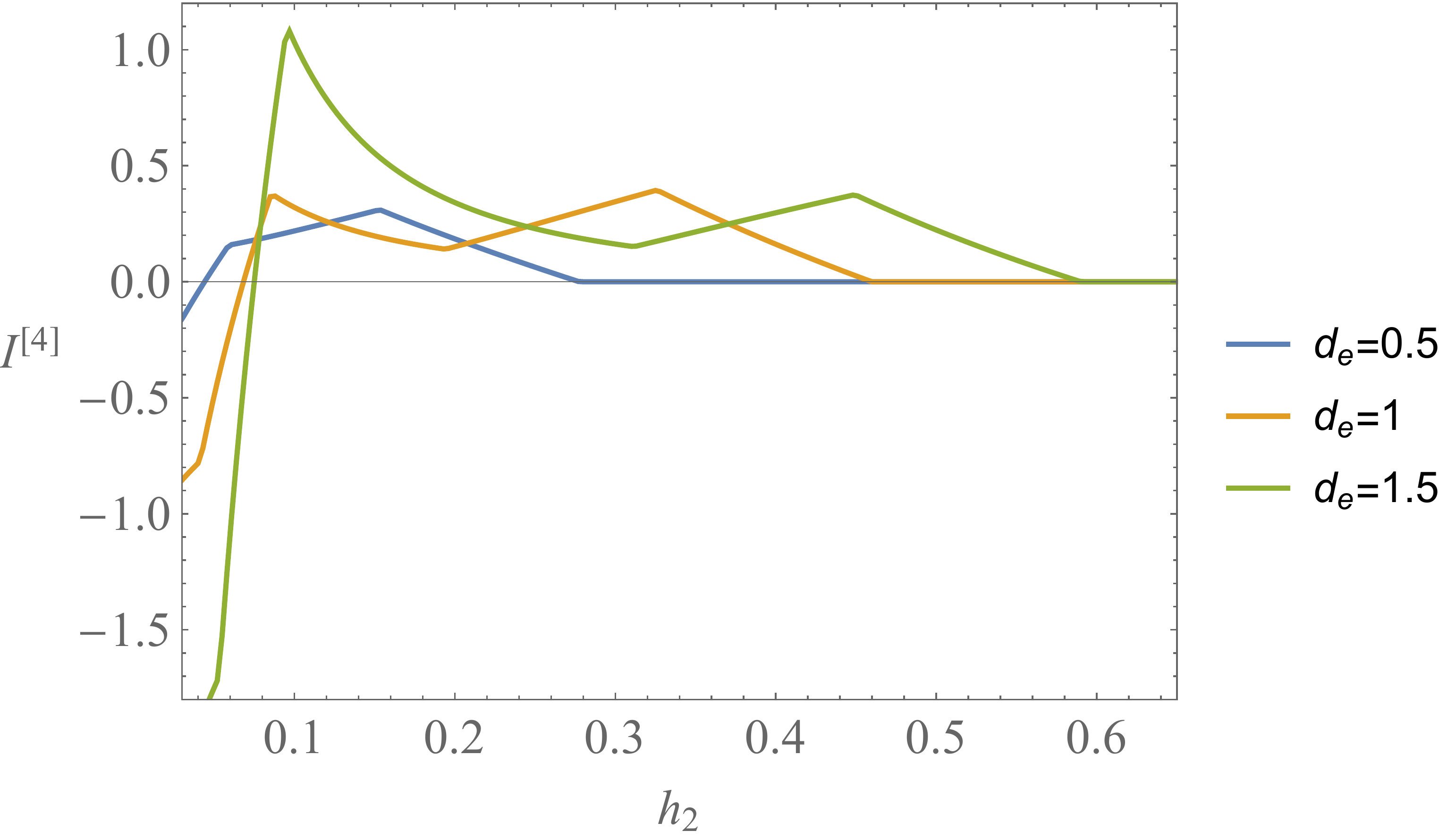}
 \caption{A counter-example of the sign property at $n=4$ in HV spacetime for different discrete values of $d_e$. The separation distance $h_2$ is set as the variable, and the strip lengths $l_i$ and separation distances $h_i$ respectively take different values.}\label{fig_counter}
\end{figure}

In section \ref{3}, we have assumed that the strips have equal length $l$ and are separated by an equal distance $h$. Such kind of simplification guarantees that holographic nI is a univariate function of $h$ by normalizing the dimensionful quantities with $l$. As usual, we have found that the conjectured sign property of holographic nI holds, in consistency with \cite{Alishahiha:2014jxa,MohammadiMozaffar:2015wnx,Mirabi:2016elb,Mahapatra:2019uql}. Counter-examples violating the conjecture were found in $\text{AdS}_4$ spacetime/black hole for $n=5$ in \cite{Mirabi:2016elb}, in $\text{AdS}_3$ spacetime for $n=4,5$ in \cite{Hayden:2011ag}, and numerically detected for static and dynamical backgrounds for $n=4,5$ in \cite{Erdmenger:2017gdk}. Now we show that counter-examples are not uncommon with the most generic configuration for HEE of the strips, i.e. arbitrary strip lengths $l_i$ and distances $h_i$. We seek counter-examples for $I^{[4]}$ in HV spacetime for convenience, with results displayed in figure \ref{fig_counter}, where we require on purpose that $h_i,l_2,l_3$ be small. In the full range of $d_e>0$, including $d_e=d$ corresponding to AdS spacetime, the conjectured sign property can be violated. 
We conclude that $I^{[3]}\leq 0$ holds strictly, that is, HMI is monogamous, and that the sign of $I^{[n]}$ is indefinite for $n\geq 4$ with the generic configuration. In particular, our conclusion is consistent with the numerical results for $n=4,\;5$ in \cite{Erdmenger:2017gdk}. In the case that nI has a definite sign, the absolute value can be regarded as the degree of entanglement. If nI has no definite sign, (1) such kind of physical interpretation becomes obscure, (2) nI must vanish (indicating no entanglement) at some point while on the two sides it is non-vanishing (indicating finite entanglement), which is also unphysical. In short, we interpret the indefinite sign of $I^{[n]}$ for $n\geq 4$ as the first evidence that it may not be a well-defined entanglement measure. The special case is that the strips are of equal length and separated by an equal distance. We find in this case that $I^{[n]}\geq 0/I^{[n]}\leq 0$ for even/odd $n$ in HV geometry.

We have also explored the monotonicity of $I^{[n]}(h)$ at small $h$. ${I^{[3]}}(h)$ is a decreasing function for arbitrary $d_e>0$. For $n>3$, ${I^{[n]}}(h)$ is an in/decreasing function at small/large $d_e$ if $n$ is even, and is the opposite if $n$ is odd. This is analysed in more detail in section \ref{3.4}. In general, the ${I^{[2]}}$ is a monotonically decreasing function of $h$ while the ${I^{[n]}}-h$ relation is non-monotonic for $n\geq 3$, considering the sign of $I^{[n]}$ and the fact that ${I^{[n]}}(h)$ undergoes only first-order phase transitions and vanishes at large $h$. We expect that ${I^{[2]}}$ correctly measures the attenuating entanglement as the disjoint strips separate, and ${I^{[n]}}$ problematic for $n\geq 3$.

Furthermore, HEE is an increasing function of $d_e$, which reflects that higher dimensions correspond to more entangling degrees of freedom in a pure state. $I^{[2]}$ also increases as $d_e$ increases while the relation becomes complicated for $n\geq 3$: roughly speaking, ${I^{[n]}}(h)$ curves are stretched towards the positive $h$-axis by increasing $d_e$. The $I^{[2]}-d_e$ relation indicates generally in a mixed state that the bipartite entanglement increases with the growth of the degrees of freedom. In contrast, the non-monotonic $I^{[n]}-d_e$ relation for larger $n$ may indicate that $I^{[n]}$ is an ill-conditioned entanglement measure.

Temperature would enhance HEE but suppress the absolute value of $I^{[n]}$. We expect that at higher temperature, the quantum correlations are strengthened while the total entangled degrees of freedom tend to get disentangled. The UV cutoff $r_c$ suppresses HEE and $I^{[2]}$, and has an effect of non-monotonically stretching the $I^{[n]}(h)$ curves towards the negative $h$-axis for larger $n$. This non-monotonic $I^{[n]}-r_c$ relation may also imply that $I^{[n]}$ is unphysical. Through analysis above, we interpret the non-monotonic $I^{[n]}-\{h,\;d_e,\;r_c\}$ relations for $n\geq 3$ as the second evidence that $I^{[n]}$ may not be a well-defined entanglement measure.

We find that $I^{[2]}$ is a convex function of $h$ before the phase transition, and in contrast that for boundary strips of equal lengths separated by equal distances, $I^{[n]}(h)$ for $n\geq 3$ is quasi-linear in each segment corresponding to different RT configurations. The $I^{[2]}-h$ relation measures the attenuating entanglement as the two strips separate, while the $I^{[n]}-h$ relation for $n\geq 3$ seems unphysical. Therefore, we interpret the quasi-linearity of $I^{[n]}-h$ for $n\geq 3$ as the third evidence that $I^{[n]}$ may not be a well-defined entanglement measure. Certainly, the $I^{[n]}-h$ relation displays stronger nonlinearity in more general configurations of the strips, e.g. figure \ref{fig_counter}, while not negligible is the quasi-linear relation in the special configuration where the strips are of equal lengths separated by equal distances. 

 A summary of the properties of the holographic entanglement measures goes as follows. HEE of one single strip was found to be a positive, UV divergent, increasing and concave function of the strip length (or more generally the scale of the entangling region); and therefore as a linear combination of HEE, HMI between two parallel strips (or more generally two entangling subregions) was found to be a non-negative, UV divergent/finite, decreasing and convex function of the distance. We believe that HMI is a well-defined entanglement measure of two disjoint subsystems, seen by definition. For $n=3$, we find $I^{[3]}$ to be non-positive, UV finite, non-monotonic in general, and quasi-linear by identifying the strip lengths and distances. The negative sign is fine since we may as well take the absolute value as the degree of tripartite entanglement. However, the non-monotonicity and quasi-linearity in the special case cast doubt on whether $I^{[3]}$ is qualified as a physical measurement. For $n>3$, even the sign of $I^{[n]}$ becomes indefinite although $I^{[n]}$ inherits the other properties at $n=3$. 

It has been a belief that the sign property of holographic nI in contrast with the indefinite sign of nI in field theories helps examine the holographic duality, i.e. only the quantum many-body theories in which nI has a definite sign could have the dual gravitional description. Now we deem with our computation that this belief is naive since holographic nI, consistent with nI, does not obey the sign property in general. Remarkably, $n=3$ is an exception in that HTI has a definite sign as proved with RT prescription, while TI does not in field theories, with the physical explanation to be found. 

As an aside, we comment on the configurations of $S(\cup_i A_i)$. In the case of two strips, the physical interpretation of the two configurations is crystal clear. The connected configuration, predominant at small distance, indicates that the two strips are entangled; the disconnected configuration, predominant at large distance, indicates that there is no entanglement. 
In the case of three or more strips, we expect that a generic configuration implicates the existence of entanglement between the strips within each connected component. For instance, the connected configuration implicates that all strips are entangled at small distance, and the maximally disconnected configuration implicates that there is no multi-partite entanglement at large distance. 
%In the case of three or more strips, such kind of interpretation is ambiguous. Only the connected configuration at small distance and the maximally disconnected configuration at large distance seem to be the natural extension of the two configurations in the case of two strips, while the other disconnected configurations are purely geometric and possibly unphysical.

The huge discrepancies in the $I^{[n]}$ behaviors at $n=2,3$ and at $n>3$ motivate us to reconsider if holographic nI (or nI) is an effective measurement of multi-partite entanglement. As the strips separate, the phase transitions are allowed in gravitational computation at leading order since they are expected to be canceled by quantum corrections \cite{Hayden:2011ag}. Therefore, we anticipate that a well-defined measure of multi-partite entanglement should obey the following conditions: it captures the entanglement between any two subsystems; it is non-negative; it is a decreasing and concave function of the separation distance; the relations to other geometric parameters consist with the HMI ($I^{[2]}$) relations. Obviously, HMI satisfies the laws above while $I^{[n]}$ does not for larger $n$. Promising candidates are multitudinous, for example, $n$-partite entanglement (nE), which we leave for future research.

Recently, multi-partite entanglement measures have been generally constructed by using the replicated multi-partite state of the system \cite{Gadde:2023zzj}. These measures are gauge invariant under the replica symmetry and the system is assumed to be in a pure state. In this respect, EE and MI for pure states are qualified as bipartite entanglement measures while nI for $n\geq 3$ should be excluded. Therefore, their viewpoint of multi-partite state construction backs up our geometric analysis of holographic nI. It would be interesting to probe the relationship between the intrinsic symmetries and the geometric behaviors of the entanglement measures in future.

In the end, we list a few directions for future studies of holographic multi-partite entanglement: (1) generalization to higher dimensions, in which one has to consider the convexity condition (we discussed this in \cite{Ju:2023bjl}), (2) generalization to large scale in pure or thermal backgrounds in global coordinates, (3) computation in other holographic models satisfying the null energy condition and the nesting rule of HEE \cite{Bousso:2022hlz}, e.g. AdS/BCFT, and (4) other candidates for multi-partite entanglement measurements, either quantum mechanical like conditional mutual information (CMI) for two subsystems, or both quantum mechanical and classical.
% \paragraph{Up to paragraphs.} We find that having more levels usually
% reduces the clarity of the article. Also, we strongly discourage the
% use of non-numbered sections (e.g.~\texttt{\textbackslash
%   subsubsection*}).  Please also consider the use of
% ``\texttt{\textbackslash texorpdfstring\{\}\{\}}'' to avoid warnings
% from the \texttt{hyperref} package when you have math in the section titles.

\acknowledgments    

We thank Hyun-Sik Jeong and Wen-Bin Pan for collaboration at an early stage of the work and their helpful suggestions. This work was supported by the National Key R\&D Program of China (Grant No. 2018FYA0305800), Project 12035016 supported by National Natural Science Foundation of China, the Strategic Priority Research Program of Chinese Academy of Sciences (Grant No. XDB28000000).

\bibliography{biblio}
\bibliographystyle{JHEP}

\end{document}